\documentclass[twocolumn,aps,prl,amsmath,showpacs,amssymb,floatfix]{revtex4}
\usepackage{tabularx}
\usepackage{bm}
\usepackage{subfigure}
\usepackage{euscript}
\usepackage{graphicx}
\usepackage{color}
\usepackage[colorlinks=true,linkcolor=blue]{hyperref}%
\usepackage{amsfonts}
\usepackage{exscale}
\usepackage{amsbsy}

\begin{document}

\title{Antiferromagnetic Order in a Spin-Orbit Coupled Bose-Einstein Condensate}

\author{Zhongbo Yan}
\author{Shaolong Wan}
\email[]{slwan@ustc.edu.cn}
\affiliation{Institute for Theoretical
Physics and Department of Modern Physics University of Science and
Technology of China, Hefei, 230026, P. R. China}

\date{\today}

\begin{abstract}
Spin-orbit coupling related new physics and quantum magnetism are
two branches of great interest both in condensed matter physics
and in cold atomic physics. With the introduction of a Rashba-like
SOC into a Bose-Einstein condensate (BEC) loaded in a
two-dimensional bipartite optical square lattice, we find that the
ground state of the BEC always favors a coherent condensate than a
fragmented condensate and always exhibits very large degeneracy,
and most importantly, an antiferromagnetic order of quantum nature
emerges when parameters satisfy certain condition. This provides
an ideal platform to study the interplay of antiferromagnetic
phase and superfluid phase.
\end{abstract}

\pacs{03.75.Mn, 67.85.Fg, 75.10.-b}

\maketitle {\it Introduction.---} Due to the great impact of
spin-orbit coupling (SOC) on the band structure of both fermionic
systems and bosonic systems, the study of new physics related to
SOC has been being of central interest both in condensed matter
community and in cold atomic community for many years \cite{M. Z.
Hasan, X.-L. Qi, V. Galitski, H. Zhai, X. F. Zhou, W. Yi, Y. Xu, S. Z. Zhang}.
For fermionic systems with certain symmetries, it is found that in the
appearance of SOC, the energy gap of bands usually gets closed and
reopened in a nontrivial way accompanying a topological phase
transition \cite{A. P. Schnyder, S. Ryu}. For bosonic systems, SOC
usually induces a shift of the energy minima from zero momentum to
nonzero momentum with a number increase of the minima, as a
result,  the ground state of a spin-orbit coupled Bose-Einstein
condensate (BEC) will have many possibilities and may exhibit new
exotic phases of great interest \cite{T. D. Stanescu, Y.-J. Lin,
C. Wang, C.-M. Jian, C. J. Wu, T.-L. Ho, S.-K. Yip, S.
Gopalakrishnan1, Yun Li, T. Ozawa, S. Gopalakrishnan2, Y. Deng, R.
M. Wilson, S.-C. Ji}.

Quantum magnetism, due to its fundamental importance in
understanding many-body physics and its great potential
applications in real life, is always one of the hottest fields in
condensed matter physics \cite{A. Auerbach, S. Sachdev}. The
simplest many-body model that exhibits quantum magnetism is the
well-known Fermi-Hubbard model which plays a crucial role in
understanding the high-$T_{c}$ superconductor, however, as
material systems always exhibit inevitable complexity, like
defects, even though extensive efforts have been put in,  a fully
understanding of this simple model seems still far away.
Therefore, to get a better understanding of the quantum magnetism
in a controllable way, recently, several groups have put much efforts
in engineering and observing magnetic order in cold atomic optical
lattice systems \cite{J. Simon, J. Struck1, D. Greif, J. Struck2,
C. V. Parker}. These systems include both fermionic ones \cite{D.
Greif} and bosonic ones \cite{J. Simon, J. Struck1, J. Struck2, C.
V. Parker}. As observing exchange-driven quantum magnetism of a
fermionic system has been hindered by the required ultralow
temperatures and entropies, currently most of the experiments are
carried out in bosonic systems. For these bosonic systems, spin is
usually mapped onto other physical quantities, like site
occupation \cite{J. Simon}, momentum \cite{C. V. Parker} or the
local phase of a BEC \cite{J. Struck1, J. Struck2}. For the last
mapping, as phase is not a quantized number, such systems simulate
classical magnetism.

In this work, we study a spin-$\frac{1}{2}$ BEC with a Rashba-like
SOC  loaded in a two-dimensional bipartite optical square lattice.
Unlike previous studies that arbitrary small SOC will shift the
energy minimum from zero momentum \cite{T. D. Stanescu, Y.-J. Lin,
C. Wang, C.-M. Jian, C. J. Wu, T.-L. Ho, S.-K. Yip, S.
Gopalakrishnan1, Yun Li, T. Ozawa, S. Gopalakrishnan2, Y. Deng, R.
M. Wilson, S.-C. Ji}, here a shift of the energy minimum occurs
only when the strength of SOC $\alpha$ reaches a critical value,
$i.e.$, $\alpha>\alpha_{c}$. Furthermore, before and after the
shift happen, the minima positions are fixed and
parameter-independent. As the minima are symmetric and located at
some special points of the Brillouin zone, there can exist some
special scattering terms with total momentum equal to the
reciprocal lattice vector ${\bf G}$, consequently, it is found
that: (i) the ground state of the system always favors a coherent
BEC instead of a fragmented BEC, (ii) with phase coherence
guaranteed, the ground state always exhibits large degeneracy even
in the appearance of interaction, (iii) with fixed parameters, all
degenerate ground states correspond to the same spin
configuration, (iv) most importantly, an {\it antiferromagnetic
order of quantum nature} emerges when $\alpha>\alpha_{c}$ and
interspin interaction is larger than intraspin interaction.

{\it Theoretical model.---}The lattice model we consider in this work
is given as
\begin{eqnarray}
H &=& H_{0} + H_{int} \nonumber\\
H_{0}&=& -\frac{t}{2} \sum_{<i,j>, \sigma}
(\hat{a}_{i\sigma}^{\dag}\hat{b}_{j\sigma} + h.c.)
-\Delta\mu\sum_{i\in A,\sigma}\hat{a}_{i\sigma}^{\dag}\hat{a}_{i\sigma}+ \nonumber\\
&&\Delta\mu \sum_{i\in B,\sigma} \hat{b}_{i\sigma}^{\dag}
\hat{b}_{i\sigma}
+ \left\{\sum_{i\in A} \left[ \frac{\alpha}{2} \hat{a}_{i,\uparrow}^{\dag}(\hat{b}_{i+\hat{x},\downarrow}-\hat{b}_{i-\hat{x},\downarrow}) \right. \right. \nonumber\\
&& \left. \left. + \frac{i\alpha}{2} \hat{a}_{i,\uparrow}^{\dag}
(\hat{b}_{i+\hat{y},\downarrow}
- \hat{b}_{i-\hat{y},\downarrow}) + h.c. \right] - \left[ \hat{a}\longleftrightarrow \hat{b} \right] \right\} \nonumber\\
H_{int}&=&\sum_{i\in A,\beta\gamma}U_{\beta\gamma,A}\hat{n}_{i\beta}\hat{n}_{i\gamma}+\sum_{i\in B,\beta\gamma}U_{\beta\gamma,B}\hat{n}_{i\beta}\hat{n}_{i\gamma}\nonumber\\
&=&\sum_{i\in A}
\left[\frac{c_{0,A}}{2}\hat{n_{i}}^{2}+\frac{c_{2,A}}{2}\hat{S}_{z,i}^{2}
\right] + \left[ A\longleftrightarrow B \right], \label{1}
\end{eqnarray}
where $t$ denotes the nearest-neighbor hopping amplitude,
$\Delta\mu$ denotes the staggered potential, and $\alpha$ denotes
the strength of spin-orbit coupling. $U_{\beta\gamma,A}$ and
$U_{\beta\gamma,B}$ denote the strength of the interaction at
sublattices $A$ and $B$, respectively. $\sigma$, $\beta$ and
$\gamma$ denote the two spin degrees $\{\uparrow,\downarrow\}$.
$n_{i\in A,\beta}=\hat{a}_{i\beta}^{\dag}\hat{a}_{i\beta}$ and
$n_{i\in B,\beta}=\hat{b}_{i\beta}^{\dag}\hat{b}_{i\beta}$ are the
particle number operators for spin $\beta$ and corresponding to
sublattice $A$ and $B$, respectively.
$\hat{n}_{i}=n_{i,\uparrow}+n_{i,\downarrow}$,
$\hat{S}_{z,i}=n_{i,\uparrow} -n_{i,\downarrow}$. Without loss of
generality, we assume $U_{\uparrow\uparrow,A(B)}
=U_{\downarrow\downarrow,A(B)}$ and we use $U_{1,A(B)}$ to denote
both of them. For $U_{\uparrow\downarrow,A(B)}$, we use
$U_{2,A(B)}$ to denote it. Based on these,
$c_{0,A(B)}=(U_{1,A(B)}+U_{2£¬A(B)})$,
$c_{2,A(B)}=(U_{1,A(B)}-U_{2,A(B)})$, and the sign of $c_{2,A}$
and $c_{2,B}$ are the same.

By using a Fourier transformation, the Hamiltonian without interaction under
the representation $\Phi_{k}=(\hat{a}_{k\uparrow},  \hat{b}_{k\downarrow},
\hat{b}_{k\uparrow}, \hat{a}_{k\downarrow})^{T}$ is given as
\begin{eqnarray}
\mathcal{H}_{0}(k)=\epsilon_{k}\tau_{x}-\Delta\mu\sigma_{z}\tau_{z}+\Lambda_{k}\tau_{z},\label{2}
\end{eqnarray}
where $\epsilon_{k}=-t(\cos(k_{x}a)+\cos(k_{y}a))$ corresponds to
the kinetic term,
$\Lambda_{k}=\alpha(\sin(k_{x}a)\sigma_{y}-\sin(k_{y}a)\sigma_{x})$
is the SOC which has a Rashba form. Note
$\{\epsilon_{k}\tau_{x},\Lambda_{k}\tau_{z}\}=0$, this is
different from the usual situation where kinetic term is
commutative with the SOC term. As we will see, this difference
induces quite different physical results. By making a
transformation of the representation: $\tilde{\Phi}_{k}=(\hat
{\alpha}_{1k},  \hat{\alpha}_{2k}, \hat{\beta}_{1k},
\hat{\beta}_{2k})^{T}=U(k)\Phi_{k}$ (see Supplementary Materials),
the Hamiltonian (\ref{2}) is diagonalized as
\begin{eqnarray}
H_{0}=\sum_{k}\left[-E_{k}(\hat{\alpha}_{1k}^{\dag}\hat{\alpha}_{1k}+\hat{\alpha}_{2k}^{\dag}\hat{\alpha}_{2k})+
E_{k}(\hat{\beta}_{1k}^{\dag}\hat{\beta}_{1k}+\hat{\beta}_{2k}^{\dag}\hat{\beta}_{2k})
\right], \label{3}
\end{eqnarray}
where
\begin{eqnarray}
E(k)=\sqrt{(\Delta\mu)^{2}+\epsilon_{k}^{2}+\alpha^{2}(\sin^{2}(k_{x}a)+\sin^{2}(k_{y}a))},\label{4}
\end{eqnarray}
the spectra have double degeneracy due to time-reversal symmetry:
$\sigma_{x}\tau_{x}\mathcal{H}_{0}(k)\tau_{x}\sigma_{x}=\mathcal{H}_{0}^{*}(-k)$.
For half-filling fermionic case with chemical potential $\mu=0$,
$\mathcal{H}_{0}(k)$ also holds particle-hole symmetry:
$\sigma_{x}\tau_{z}\mathcal{H}_{0}(k)\tau_{z}\sigma_{x}=-\mathcal{H}_{0}^{*}(-k)$,
and chiral symmetry:
$\tau_{y}\mathcal{H}_{0}(k)\tau_{y}=-\mathcal{H}_{0}(k)$,
therefore, it belongs to the BDI-class \cite{A. P. Schnyder, S.
Ryu, A. Y. Kitaev}. Such spin-orbit coupled system in one
dimension under certain condition can exhibit nontrivial
topological properties \cite{Z. B. Yan}. However, in this work we
focus on a bosonic system where $\mu$ is always nonzero, what we
concern is the lower band's minima where the bosons will be
condensed, instead of the band gap in the fermionic case.

\begin{figure}
\subfigure{\includegraphics[width=4cm, height=3cm]{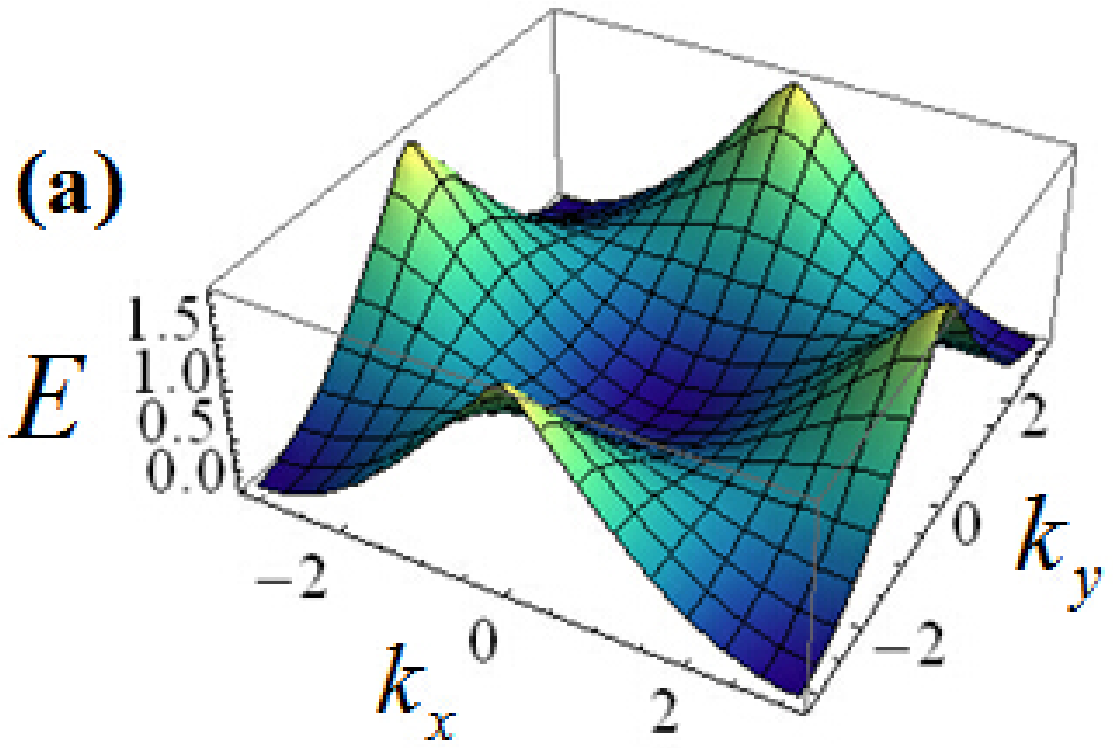}}
\subfigure{\includegraphics[width=4cm, height=3cm]{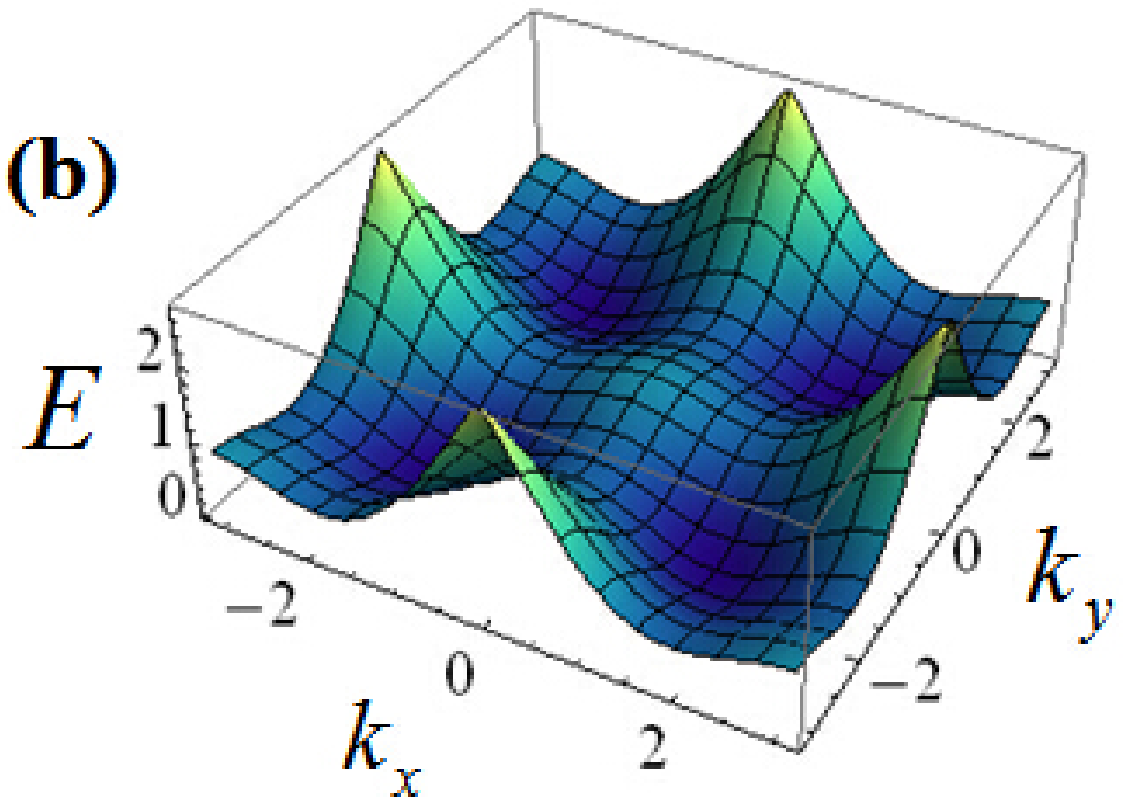}}
\subfigure{\includegraphics[width=4cm, height=4cm]{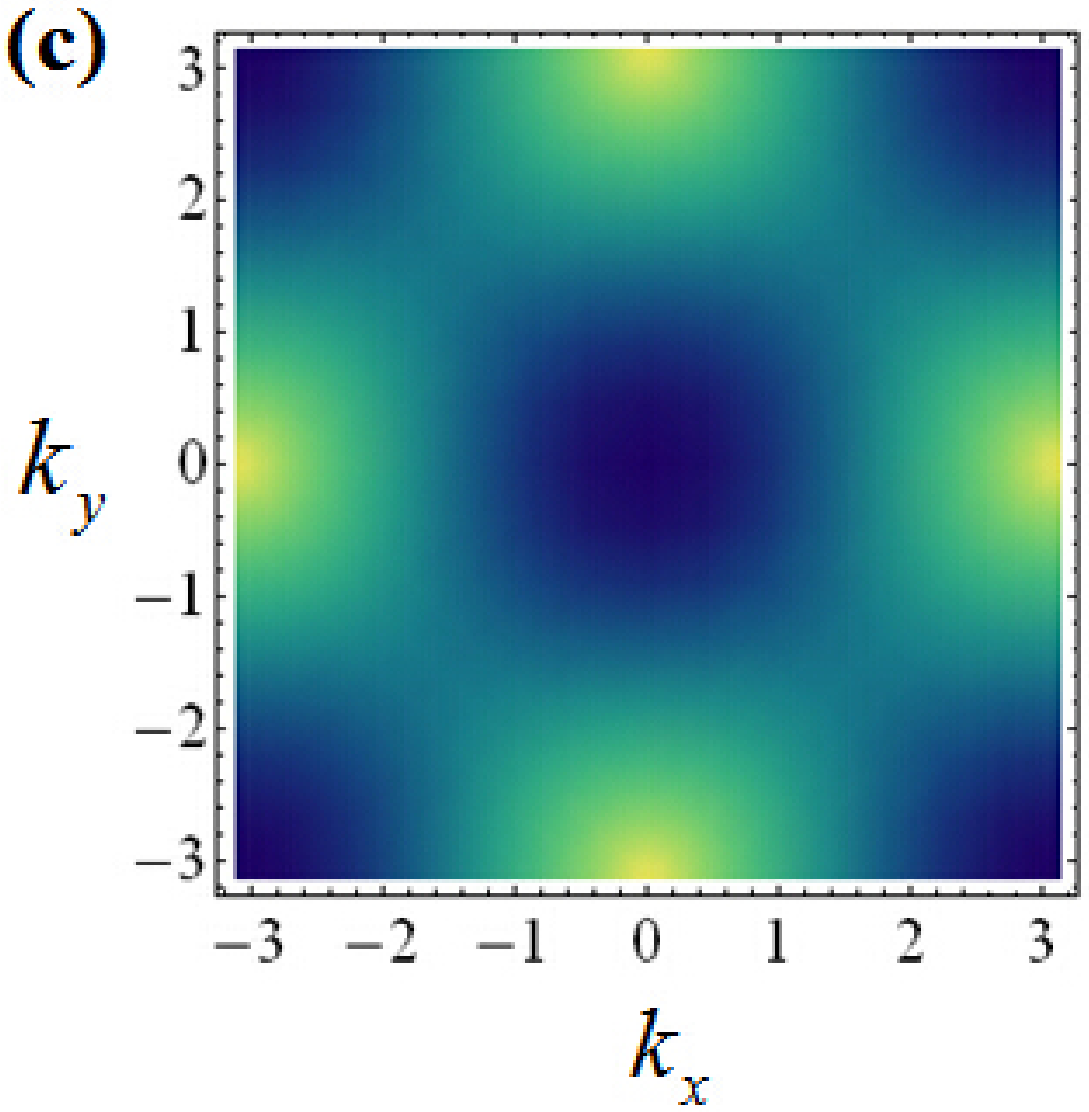}}
\subfigure{\includegraphics[width=4cm, height=4cm]{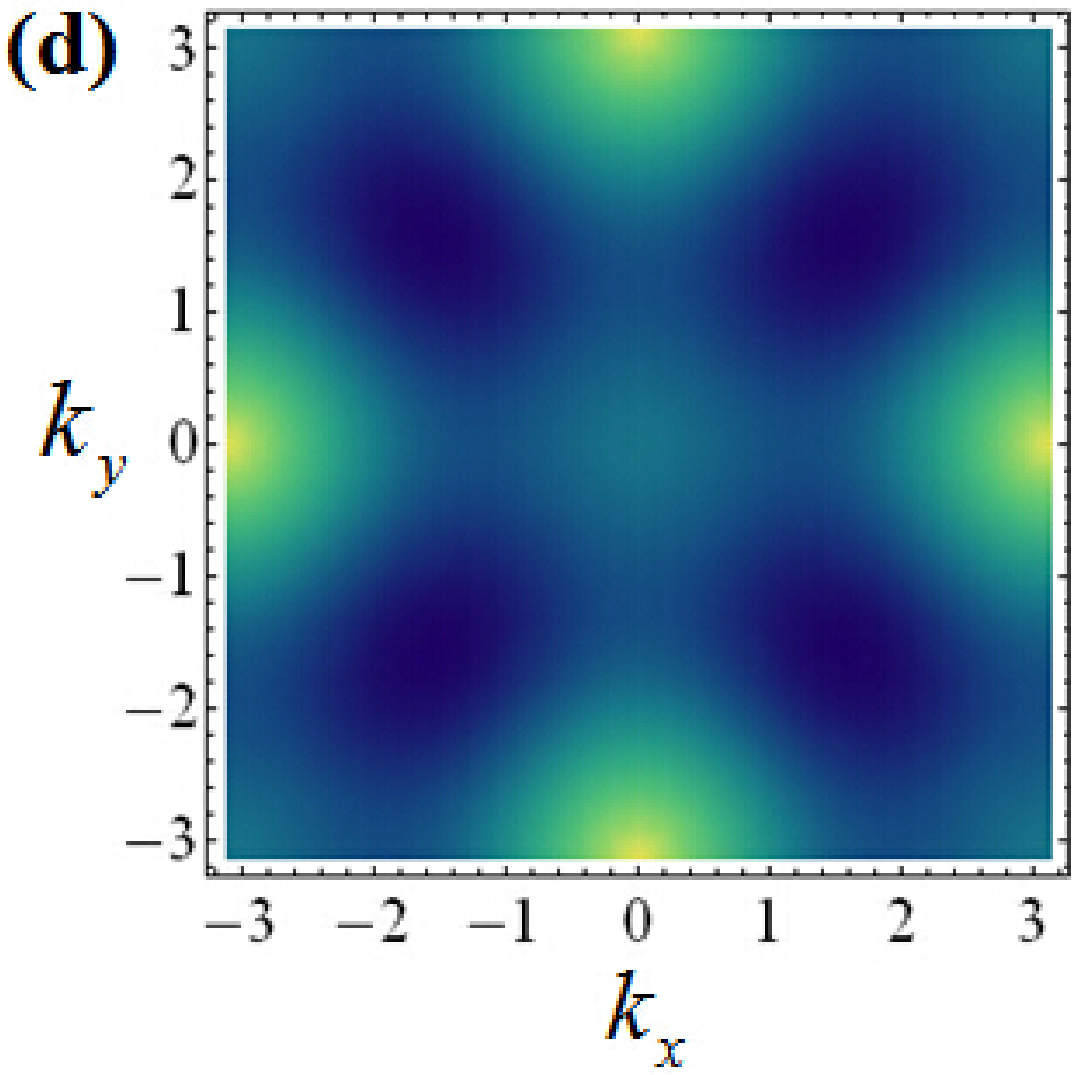}}
\caption{ (color online) (a)(b) Energy spectra correspond to
Eq.(\ref{4}). Parameters: we set $t=1$ as the energy unit, and
$a=1$ as the length unit. (a)(c) $\Delta\mu=0.1$,
$\alpha=1<\alpha_{c}$. (b)(d) $\Delta\mu=0.1$,
$\alpha=2>\alpha_{c}$. (c) is the density plot of (a), and (d) is
the density plot of (b). The deep color regions correspond to the
energy minima. }\label{fig1}
\end{figure}

For bosons at low temperature, they will be condensed at the
energy minima (we consider $T=0$ in this work). Usually, there is
only one minimum which is located at ${\bf k_{0}}=0$. However,
from Eq.(\ref{4}) or more directly from Fig.\ref{fig1}, it is
found that for this model, when $\alpha<\sqrt{2}t$, there are two
minima which are stably located at ${\bf k_{0}}=0$ and ${\bf
k_{\pi}}=(\pi/a,\pi/a)$, and when $\alpha>\sqrt{2}t$, there are
four degenerate minima stably located at ${\bf
Q_{1}}=(\pi/2a,\pi/2a)$, ${\bf Q_{2}}=(-\pi/2a,-\pi/2a)$, ${\bf
Q_{3}}=(\pi/2a,-\pi/2a)$, ${\bf Q_{4}} =(-\pi/2a,\pi/2a)$.
Increasing the strength of SOC $\alpha$ across the critical value
$\alpha_{c}=\sqrt{2}t$, the minimum where the bosons are condensed
will be shifted. As we will see, the shift is nontrivial,  it not
only directly alters the ground states, but also can establish an
{\it antiferromagnetic order} in the condensate.

{\it $\alpha<\alpha_{c}$, two minima case.---} When
$\alpha<\alpha_{c}$, to determine which minimum the bosons are
condensed at, we introduce the wave functions which correspond to
a fragmented and a coherent condensate, respectively, as \cite{C.
J. Wu}
\begin{eqnarray}
&&|\Psi_{f}>=\frac{1}{\sqrt{M}}(\alpha_{1{\bf k_{0}}}^{\dag})^{N_{10}}(\alpha_{1{\bf k_{\pi}}}^{\dag})^{N_{1\pi}}
(\alpha_{2{\bf k_{0}}}^{\dag})^{N_{20}}(\alpha_{2{\bf k_{\pi}}}^{\dag})^{N_{2\pi}}|0>,\nonumber\\
&&|\Psi_{c}>=\frac{1}{\sqrt{\Omega}}\{\lambda_{1}\hat{\alpha}_{1{\bf k_{0}}}^{\dag}+
e^{i\phi_{1}}\lambda_{2}\hat{\alpha}_{1{\bf k_{\pi}}}^{\dag}\}^{N_{1}}\nonumber\\
&&\qquad\qquad\{\lambda_{3}\hat{\alpha}_{2{\bf k_{0}}}^{\dag}+
e^{i\phi_{2}}\lambda_{4}\hat{\alpha}_{2{\bf k_{\pi}}}^{\dag}\}^{N_{2}}|0>,\label{5}
\end{eqnarray}
where $M=N_{10}!N_{1\pi}!N_{20}!N_{2\pi}!$, $\lambda_{1}=\sqrt{N_{10}/N_{1}}$,
$\lambda_{2}=\sqrt{N_{1\pi}/N_{1}}$, $\lambda_{3}=\sqrt{N_{20}/N_{2}}$,
$\lambda_{4}=\sqrt{N_{2\pi}/N_{2}}$, $\Omega=N_{1}!N_{2}!$. The particle number
partition $(N_{10},N_{1\pi})$ and $(N_{20},N_{2\pi})$ satisfy
the constraint: $N_{10}+N_{1\pi}=N_{1}$,
$N_{20}+N_{2\pi}=N_{2}$. Without loss of
generality, we assume $N_{1}=N_{2}=N/2$ where
$N$ is the total particle number in the condensate.

Since the kinetic energy of a condensate is negligible, we only
need to consider the interaction energy $<\Psi|H_{int}|\Psi>$.
Based on the wave function of a fragmented condensate, the
expression is given as \cite{P. Nozieres}
\begin{eqnarray}
&&<\Psi_{f}|H_{int}|\Psi_{f}>=[U_{1,A}N_{1}(N_{1}-1)+U_{1,A}N_{2}(N_{2}-1)\nonumber\\
&& \qquad\qquad+2U_{2,A}N_{1}N_{2}]\cos^{4}(\theta/2)+[U_{1,B}N_{1}(N_{1}-1) \nonumber\\
&&\qquad\qquad+U_{1,B}N_{2}(N_{2}-1)+2U_{2,B}N_{1}N_{2}]\sin^{4}(\theta/2) \nonumber\\
&& \qquad\qquad+2[U_{1,A}(N_{10}N_{1\pi}+N_{20}N_{2\pi})\cos^{4}(\theta/2)\nonumber\\
&& \qquad\qquad+U_{1,B}(N_{10}N_{1\pi}+N_{20}N_{2\pi})\sin^{4}(\theta/2)]\nonumber\\
&& \qquad\qquad=E_{s}+E_{Fock},\label{6}
\end{eqnarray}
where $E_{s}$ is the sum of the terms in the first three lines,
and $E_{Fock}$ is the sum of the terms in the fourth and fifth
lines. Due to the Fock terms, the fragmented condensate, compared
to a {\it single condensate} (either $N_{0}=0$ or $N_{\pi}=0$)
which only has energy $E_{s}$, always costs more energy and
therefore is unfavored. Based on the wave function of a coherent
condensate,
\begin{eqnarray}
&&<\Psi_{c}|H_{int}|\Psi_{c}>=<\Psi_{f}|H_{int}|\Psi_{f}>+2[U_{1,A}N_{10}N_{1\pi}\nonumber\\
&&\qquad \times \frac{(N_{1}-1)}{N_{1}}\cos(2\phi_{1})+U_{1,A}N_{20}N_{2\pi}\frac{(N_{2}-1)}{N_{2}}\cos(2\phi_{2})\nonumber\\
&&\qquad +4U_{2,A}\sqrt{N_{10}N_{1\pi}N_{20}N_{2\pi}}\cos(\phi_{1})\cos(\phi_{2})]\cos^{4}(\theta/2)\nonumber\\
&&\qquad +2[U_{1,B}N_{10}N_{1\pi}\frac{(N_{1}-1)}{N_{1}}\cos(2\phi_{1})+U_{2,B}N_{20}N_{2\pi}\nonumber\\
&&\qquad \times\frac{(N_{2}-1)}{N_{2}}\cos(2\phi_{2})+4U_{2,B}\sqrt{N_{10}N_{1\pi}N_{20}N_{2\pi}}\nonumber\\
&&\qquad\times\cos(\phi_{1})\cos(\phi_{2})]\sin^{4}(\theta/2),\label{7}
\end{eqnarray}
where $\theta=\arctan(2t/\Delta\mu)$. Compared
$<\Psi_{c}|H_{int}|\Psi_{c}>$ to $<\Psi_{f}|H_{int}|\Psi_{f}>$,
the additional terms appearing in Eq.(\ref{7}) is due to the fact
that the system is a lattice one, therefore, unlike the continue
case, such terms like $\hat{\alpha}_{1{\bf k_{\pi}}}^{\dag}
\hat{\alpha}_{1{\bf k_{\pi}}}^{\dag}\hat{\alpha}_{1{\bf k_{0}}}
\hat{\alpha}_{1{\bf k_{0}}}$ and $\hat{\alpha}_{1{\bf
k_{\pi}}}^{\dag} \hat{\alpha}_{2{\bf
k_{\pi}}}^{\dag}\hat{\alpha}_{2{\bf k_{0}}} \hat{\alpha}_{1{\bf
k_{0}}}$ are allowed because $2({\bf k_{\pi}} -{\bf k_{0}})={\bf
G}$, where ${\bf G}$ is the reciprocal vector. The appearance of
these additional terms makes the {\it coherent condensate always
more favored than the fragmented condensate} since
$<\Psi_{c}|H_{int}|\Psi_{c}>$ can always be made to be smaller
than $<\Psi_{f}|H_{int}|\Psi_{f}>$ by tuning the phase $\phi_{1}$
and $\phi_{2}$. Therefore, to determine the ground state, what we
need to do is to minimize $<\Psi_{c}|H_{int}|\Psi_{c}>$.

As $N$ is generally large,  $(N_{i}-1)/N_{i}$ can be taken as $1$.
It is found that when $U_{2}<U_{1}$,  $<\Psi_{c}|H_{int}|\Psi_{c}>$
takes the same minimum value $E_{s}$ for arbitrary particle number
partition if the phases $\phi_{1}$ and $\phi_{2}$ are locked to
$\{(n+\frac{1}{2})\pi, n\in Z\}$. Therefore, the degeneracy
of the ground state is very large ($\propto N^{2}$), these degenerate
ground states can be written compactly as
\begin{eqnarray}
|\Psi_{g}>&=&\frac{1}{\sqrt{\Omega}}\{\lambda_{1}\hat{\alpha}_{1{\bf k_{0}}}^{\dag}\pm i
\lambda_{2}\hat{\alpha}_{1{\bf k_{\pi}}}^{\dag}\}^{N_{1}}\nonumber\\
&&\{\lambda_{3}\hat{\alpha}_{2{\bf k_{0}}}^{\dag}
\pm i\lambda_{4}\hat{\alpha}_{2{\bf k_{\pi}}}^{\dag}\}^{N_{2}}|0>.\label{8}
\end{eqnarray}
When $U_{2}>U_{1}$, the ground state wave function keeps its form
in Eq.(\ref{8}), but to reach the ground state, the system will
undergo a phase separation. Besides, there emerges two new
possible ground states where $N_{10}=N_{1\pi}=N_{20}=N_{2\pi}=N/4$
and $\phi_{1}$ and $\phi_{2}$ are either given as $\phi_{1}=0$ and
$\phi_{2}=\pi$ or $\phi_{1}=\pi$ and $\phi_{2}=0$. The two phases
turn out to be locked to each other. The two new possible ground
state wave functions can be written as
\begin{eqnarray}
|\tilde{\Psi}_{g}>=\frac{1}{\sqrt{2^{N}\Omega}}\{\hat{\alpha}_{1{\bf
k_{0}}}^{\dag}\pm \hat{\alpha}_{1{\bf
k_{\pi}}}^{\dag}\}^{N_{1}}\{\hat{\alpha}_{2{\bf k_{0}}}^{\dag} \mp
\hat{\alpha}_{2{\bf k_{\pi}}}^{\dag}\}^{N_{2}}|0>. \label{9}
\end{eqnarray}

\begin{figure}
\subfigure{\includegraphics[width=8cm, height=7cm]{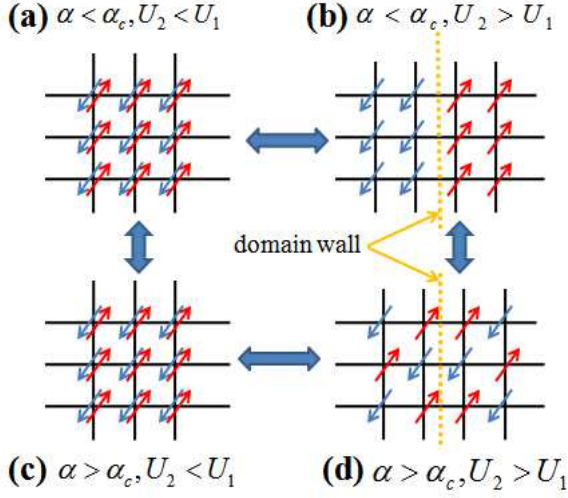}}
\caption{ (color online) Spin configurations correspond to different parameter regions.
(a)(c) paramagnetic configuration, (b) ferromagnetic configuration, (d) antiferromagnetic
configuration. The domain walls in (b)(d) are a result of
number conservation for each spin.}\label{fig2}
\end{figure}

Although the ground state has very large degeneracy, the system in
real space will only exhibit two kinds of spin-configurations. In
order to show this, we write down the spinor wave function
corresponding to the condensate in real space,
\begin{eqnarray}
\vec{\varphi}({\bf r})=
\sqrt{n_{0}}[(a_{1}\vec{e}_{1}+a_{2}\vec{e}_{2})e^{i{\bf k_{\pi}\cdot r}}+(a_{3}\vec{e}_{3}+a_{4}\vec{e}_{4})],\label{10}
\end{eqnarray}
where $n_{0}=N/N_{T}$ is the condensation density with $N_{T}$ the
number of lattice sites,
$\vec{\varphi}=[\varphi_{A,\uparrow},\varphi_{B,\downarrow},\varphi_{B,\uparrow},\varphi_{A,\downarrow}]^{T}$,
$\vec{e}_{1}=[\chi_{1},0,-\chi_{2},0]^{T}$,
$\vec{e}_{2}=[0,-\chi_{2},0,\chi_{1}]^{T}$,
$\vec{e}_{3}=[\chi_{1},0,\chi_{2},0]^{T}$.
$\vec{e}_{4}=[0,\chi_{2},0,\chi_{1}]^{T}$, with
$\chi_{1}=\cos(\theta/2)$, $\chi_{2}=\sin(\theta/2)$,
$\theta=\arctan(2t/\Delta\mu)$. $a_{i}$ are complex coefficients
which satisfy $|a_{1}|^{2}+|a_{3}|^{2}=|a_{2}|^{2}+|a_{4}|^{2}=1$
and are determined by minimizing the energy functional
\begin{eqnarray}
\mathcal{\varepsilon}&=&\sum_{i\in A}[\frac{c_{0,A}}{2}(|\varphi_{i,\uparrow}|^{2}+|\varphi_{i,\downarrow}|^{2})^{2}
+\frac{c_{2,A}}{2}(|\varphi_{i,\uparrow}|^{2}-|\varphi_{i,\downarrow}|^{2})^{2}]\nonumber\\
&&\qquad+(A\longleftrightarrow B).\label{11}
\end{eqnarray}
From Eq.(\ref{10}), it is direct to obtain
\begin{eqnarray}
&&|\varphi_{i\in A,\uparrow}|^{2}=n_{0}\chi_{1}^{2}(1+2|a_{1}||a_{3}|\cos[\frac{\pi(x_{i}+y_{i})}{a}+\phi_{1}])\xi_{1},\nonumber\\
&&|\varphi_{i\in B,\uparrow}|^{2}=n_{0}\chi_{2}^{2}(1-2|a_{1}||a_{3}|\cos[\frac{\pi(x_{i}+y_{i})}{a}+\phi_{1}])\xi_{2},\nonumber\\
&&|\varphi_{i\in A,\downarrow}|^{2}=n_{0}\chi_{1}^{2}(1+2|a_{2}||a_{4}|\cos[\frac{\pi(x_{i}+y_{i})}{a}+\phi_{2}])\xi_{1},\nonumber\\
&&|\varphi_{i\in B,\downarrow}|^{2}=n_{0}\chi_{2}^{2}(1-2|a_{2}||a_{4}|\cos[\frac{\pi(x_{i}+y_{i})}{a}+\phi_{2}])\xi_{2},\qquad\label{12}
\end{eqnarray}
where $\xi_{1}=[(-1)^{x_{i}/a}+(-1)^{y_{i}/a}]^{2}/4$,
$\xi_{2}=[(-1)^{x_{i}/a}-(-1)^{y_{i}/a}]^{2}/4$, here we have made
a choice that sublattices A correspond to that $x_{i}/a$ and
$y_{i}/a$ are simultaneously even or odd. When $U_{2}<U_{1}$,
$i.e.$, $c_{2,A(B)}>0$, it is not hard to obtain that when
$|\varphi_{i\in A,\uparrow}|^{2}= |\varphi_{i\in
A,\downarrow}|^{2}=n_{0}\chi_{1}^{2}\xi_{1}$, $|\varphi_{i\in
B,\uparrow}|^{2}= |\varphi_{i\in
B,\downarrow}|^{2}=n_{0}\chi_{2}^{2}\xi_{2}$, the energy
functional (\ref{11}) take its minimum value. The above condition
is satisfied for arbitrary $a_{i}$ if $\phi_{1}$ and $\phi_{2}$
are locked to $\{(n+\frac{1}{2})\pi, n\in Z\}$. We can
find that the same conclusion as the one above Eq.(\ref{8}) is
reached. Therefore, these degenerate ground states described by
Eq.(\ref{8}) all correspond to a {\it spin-balanced} or {\it
paramagnetic} condensate, shown in Fig.\ref{fig2}(a). When
$U_{2}>U_{1}$, $c_{2,A(B)}<0$, it is found that the case with
$a_{1}=a_{2}=a_{3}=a_{4}=1/2$, which corresponds to
$N_{10}=N_{1\pi}=N_{20}=N_{2\pi}=N/2$, the phases should be given
as \{$\phi_{1}(0<x_{i}<L_{x}/2,y_{i})=0$ or $\pi$,
$\phi_{2}(0<x_{i}<L_{x}/2,y_{i})=\pi$ or $0$,
$\delta\phi=\pi\mod2\pi$\} and
\{$\phi_{1}(L_{x}/2<x_{i}<L_{x},y_{i})=\pi-\phi_{1}(0<x_{i}<L_{x}/2,y_{i})$,
$\phi_{2}(L_{x}/2<x_{i}<L_{x},y_{i})=\pi-\phi_{2}(0<x_{i}<L_{x}/2,y_{i})$\}
(we have assumed $L_{y}>L_{x}$. Other phase configurations always
exhibit more stronger suppression of hopping, and therefore, are
not favored in energy). From Eq.(\ref{12}), we can obtain that
this phase configuration corresponds to a phase-separation {\it
ferromagnetic} condensate, shown in Fig.\ref{fig2}(b). Therefore,
when $U_{2}>U_{1}$, all degenerate ground states corresponds to a
{\it ferromagnetic} condensate.

{\it $\alpha>\alpha_{c}$, four minima case.---} This case is our most interested case.
When $\alpha>\alpha_{c}$, there are four energy minima ${\bf Q_{1,2,3,4}}$.  Similar
to the two minima case, we introduce two wave functions which correspond to a fragmented
and a coherent condensate, respectively, as
\begin{eqnarray}
|\tilde{\Psi}_{f}>&=&\frac{1}{\sqrt{\tilde{M}}}(\alpha_{1{\bf Q_{1}}}^{\dag})^{N_{11}}(\alpha_{1{\bf Q_{2}}}^{\dag})^{N_{12}}
(\alpha_{1{\bf Q_{3}}}^{\dag})^{N_{13}}(\alpha_{1{\bf Q_{4}}}^{\dag})^{N_{14}}\nonumber\\
&&(\alpha_{2{\bf Q_{1}}}^{\dag})^{N_{21}}(\alpha_{2{\bf Q_{2}}}^{\dag})^{N_{22}}
(\alpha_{2{\bf Q_{3}}}^{\dag})^{N_{23}}(\alpha_{2{\bf Q_{4}}}^{\dag})^{N_{24}}|0>,\nonumber\\
|\tilde{\Psi}_{c}>&=&\frac{1}{\sqrt{\Omega}}\{\lambda_{11}\alpha_{1{\bf Q_{1}}}^{\dag}
+\lambda_{12}e^{i\varphi_{1,2}}\alpha_{1{\bf Q_{2}}}^{\dag}
+\lambda_{13}e^{i\varphi_{1,3}}\alpha_{1{\bf Q_{3}}}^{\dag}\nonumber\\
&&+\lambda_{14}e^{i\varphi_{1,4}}\alpha_{1{\bf Q_{4}}}^{\dag}\}^{N_{1}}\{\lambda_{21}\alpha_{2{\bf Q_{1}}}^{\dag}
+\lambda_{22}e^{i\varphi_{2,2}}\alpha_{2{\bf Q_{2}}}^{\dag}\nonumber\\
&&+\lambda_{23}e^{i\varphi_{2,3}}\alpha_{2{\bf Q_{3}}}^{\dag}
+\lambda_{24}e^{i\varphi_{2,4}}\alpha_{2{\bf Q_{4}}}^{\dag}\}^{N_{2}}|0>,\label{13}
\end{eqnarray}
where $\tilde{M}=\prod_{i=1}^{4}N_{1i}!N_{2i}!$, $\lambda_{ij}=\sqrt{N_{ij}/N_{i}}$.
Similarly, it is direct to obtain that
\begin{eqnarray}
&&<\tilde{\Psi}_{f}|H_{int}|\tilde{\Psi}_{f}>=[U_{1,A}N_{1}(N_{1}-1)+U_{1,A}N_{2}(N_{2}-1)\nonumber\\
&& \qquad+2U_{2,A}N_{1}N_{2}]\cos^{4}(\tilde{\theta}/2)+[U_{1,B}N_{1}(N_{1}-1) \nonumber\\
&&\qquad+U_{1,B}N_{2}(N_{2}-1)+2U_{2,B}N_{1}N_{2}]\sin^{4}(\tilde{\theta}/2)+\nonumber\\
&&\qquad2\sum_{i=1< j}^{4}[U_{1,A}N_{1i}N_{1j}\cos^{4}(\frac{\tilde{\theta}}{2})+U_{1,B}N_{2i}N_{2j}
\sin^{4}(\frac{\tilde{\theta}}{2})]\nonumber\\
&&\qquad=\tilde{E}_{s}+\tilde{E}_{Fock},\label{14}
\end{eqnarray}
where the terms in the fourth line correspond to $\tilde{E}_{Fock}$, and
$\tilde{\theta}=\arctan(\sqrt{2}\alpha/\Delta\mu)$. As $\tilde{\theta}>\theta$
and $U_{A}>U_{B}$ (sublattices $B$ will be shallower than $A$), it is not hard
to see that $\tilde{E}_{s}$ is smaller than $E_{s}$, however, this decrease is
quite small at the neighborhood of the critical point, $\alpha=\alpha_{c}$, if
$N_{1j}=N_{1}/4$ and  $N_{2j}=N_{2}/4$ for arbitrary $j$, $\tilde{E}_{Fock}$
is approximately equal to $3E_{Fock}/2$. This suggests that a more fragmented
condensate costs more energy.

The concrete form of $<\tilde{\Psi}_{c}|H_{int}|\tilde{\Psi}_{c}>$
is very tedious and is given explicitly in the Supplementary
Materials. Based on $<\tilde{\Psi}_{c}|H_{int}|\tilde{\Psi}_{c}>$,
it is also found that the ground state has large degeneracy. When
$U_{2}<U_{1}$, the bosons can be condensed at: one of the minima
${\bf Q_{i}}$, or two time-reversal-partner minima $\{{\bf
Q_{i}},-{\bf Q_{i}}\}$, or four minima simultaneously with
$N_{1j}=N_{1}/4$ and  $N_{2j}=N_{2}/4$ for arbitrary $j$. The
first two cases can be described by a wave function similar to
Eq.(\ref{8}), for the last one, the ground state wave function is
given as
\begin{eqnarray}
|\tilde{\Psi}_{g}>&=&\frac{1}{\sqrt{4^{N}\Omega}}\{(\hat{\alpha}_{1{\bf Q_{1}}}^{\dag}-
\hat{\alpha}_{1{\bf Q_{2}}}^{\dag})\pm i(\hat{\alpha}_{1{\bf Q_{3}}}^{\dag}+
\hat{\alpha}_{1{\bf Q_{4}}}^{\dag})\}^{N_{1}}\nonumber\\
&&\{(\hat{\alpha}_{2{\bf Q_{1}}}^{\dag}-
\hat{\alpha}_{2{\bf Q_{2}}}^{\dag})\pm i(\hat{\alpha}_{2{\bf Q_{3}}}^{\dag}+
\hat{\alpha}_{2{\bf Q_{4}}}^{\dag})\}^{N_{2}}|0>.\label{15}
\end{eqnarray}

When $U_{2}>U_{1}$, it is found that the degeneracy is greatly reduced,
there are only eight possible degenerate ground states which can be
written compactly as
\begin{eqnarray}
|\tilde{\Psi}_{g}>&=&\frac{1}{\sqrt{2^{N}\Omega}}\{\gamma_{1}(\hat{\alpha}_{1{\bf Q_{1}}}^{\dag}\pm
\hat{\alpha}_{1{\bf Q_{2}}}^{\dag})+\gamma_{2}(\hat{\alpha}_{1{\bf Q_{3}}}^{\dag}\pm
\hat{\alpha}_{1{\bf Q_{4}}}^{\dag})\}^{N_{1}}\nonumber\\
&&\{\gamma_{3}(\hat{\alpha}_{2{\bf Q_{1}}}^{\dag}\mp
\hat{\alpha}_{2{\bf Q_{2}}}^{\dag})+\gamma_{4}(\hat{\alpha}_{2{\bf Q_{3}}}^{\dag}\mp
\hat{\alpha}_{2{\bf Q_{4}}}^{\dag})\}^{N_{2}}|0>.\label{16}
\end{eqnarray}
where $\gamma_{1,2,3,4}$ only take two values $\{0,1\}$ and $\gamma_{2}=1-\gamma_{1}$,
$\gamma_{4}=1-\gamma_{3}$. Eq.(\ref{16}) suggests that for each degree, $\hat{\alpha}_{1}$
or $\hat{\alpha}_{2}$, the bosons can only choose one pair of the time-reversal-partner
minima to condense.

Similar to the two minima case, although the degeneracy of the ground states
are large, there are also only two kinds of spin configurations. Following
the previous procedures, we first write down the spinor wave function corresponding
to the condensate,
\begin{eqnarray}
\vec{\varphi}({\bf r})&=&
\sqrt{n_{0}}[(b_{1}\vec{u}_{1}+b_{2}\vec{u}_{2})e^{i{\bf Q_{1}\cdot r}}+(b_{3}\vec{u}_{3}+b_{4}\vec{u}_{4})e^{i{\bf Q_{2}\cdot r}}\nonumber\\
&&+(b_{5}\vec{u}_{5}+b_{6}\vec{u}_{6})e^{i{\bf Q_{3}\cdot r}}+(b_{7}\vec{u}_{7}+b_{8}\vec{u}_{8})e^{i{\bf Q_{4}\cdot r}}],\label{17}
\end{eqnarray}
where $\vec{u}_{1}=[\tilde{\chi}_{1},\tilde{\chi}_{2}e^{-i\pi/4},0,0]^{T}$,
$\vec{u}_{2}=[0,0,-\tilde{\chi}_{2}e^{i\pi/4},\tilde{\chi}_{1}]^{T}$,
$\vec{u}_{3}=[\tilde{\chi}_{1},-\tilde{\chi}_{2}e^{-i\pi/4},0,0]^{T}$,
$\vec{u}_{4}=[0,0,\tilde{\chi}_{2}e^{i\pi/4},\tilde{\chi}_{1}]^{T}$,
$\vec{u}_{5}=[\tilde{\chi}_{1},-\tilde{\chi}_{2}e^{i\pi/4},0,0]^{T}$,
$\vec{u}_{6}=[0,0,\tilde{\chi}_{2}e^{-i\pi/4},\tilde{\chi}_{1}]^{T}$,
$\vec{u}_{7}=[\tilde{\chi}_{1},\tilde{\chi}_{2}e^{i\pi/4},0,0]^{T}$,
$\vec{u}_{8}=[0,0,-\tilde{\chi}_{2}e^{-i\pi/4},\tilde{\chi}_{1}]^{T}$,
with $\tilde{\chi}_{1}=\cos(\tilde{\theta}/2)$,
$\tilde{\chi}_{2}=\sin(\tilde{\theta}/2)$.
$b_{i}$ are complex coefficients which satisfy
$|b_{1}|^{2}+|b_{3}|^{2}+|b_{5}|^{2}+|b_{7}|^{2}
=|b_{2}|^{2}+|b_{4}|^{2}+|b_{6}|^{2}+|b_{8}|^{2}=1$
and are determined by minimizing Eq.(\ref{11}).

Based on Eq.(\ref{17}), the particle distribution can be directly
obtained (see Supplementary Materials). It is direct to find  that
when $U_{2}<U_{1}$, the degenerate ground states for the
aforementioned three cases all correspond to a {\it paramagnetic}
condensate, $i.e.$, $|\varphi_{i\in A,\uparrow}|^{2}=
|\varphi_{i\in A,\downarrow}|^{2}=n_{0}\chi_{1}^{2}\xi_{1}$,
$|\varphi_{i\in B,\uparrow}|^{2}= |\varphi_{i\in
B,\downarrow}|^{2}=n_{0}\chi_{2}^{2}\xi_{2}$, shown in
Fig.\ref{fig2}(c). When $U_{2}>U_{1}$, with the constraint of
particle number conservation, it is found that the particle
distribution corresponding to the ground states given in
Eq.(\ref{16}) should be: \{$|\varphi_{i\in
A\uparrow}|^{2}=2n_{0}\tilde{\chi}_{1}^{2}\xi_{1}$,
$|\varphi_{i\in A\downarrow}|^{2}=0$, $|\varphi_{i\in
B\downarrow}|^{2}=2n_{0}\tilde{\chi}_{2}^{2}\xi_{2}$,
$|\varphi_{i\in B\uparrow}|^{2}=0$, and $|\varphi_{j\in
A\uparrow}|^{2}=0$, $|\varphi_{j\in
A\downarrow}|^{2}=2n_{0}\tilde{\chi}_{1}^{2}\xi_{1}$,
$|\varphi_{j\in B\downarrow}|^{2}=0$, $|\varphi_{j\in
B\uparrow}|^{2}=2n_{0}\tilde{\chi}_{2}^{2}\xi_{2}$\}, or
\{$|\varphi_{i\in A\uparrow}|^{2}=0$, $|\varphi_{i\in
A\downarrow}|^{2}=2n_{0}\tilde{\chi}_{1}^{2}\xi_{1}$,
$|\varphi_{i\in B\downarrow}|^{2}=0$, $|\varphi_{i\in
B\uparrow}|^{2}=2n_{0}\tilde{\chi}_{2}^{2}\xi_{2}$, and
$|\varphi_{j\in
A\uparrow}|^{2}=2n_{0}\tilde{\chi}_{1}^{2}\xi_{1}$,
$|\varphi_{j\in A\downarrow}|^{2}=0$, $|\varphi_{j\in
B\downarrow}|^{2}=2n_{0}\tilde{\chi}_{2}^{2}\xi_{2}$,
$|\varphi_{j\in B\uparrow}|^{2}=0$\}, with $0<x_{i}<L_{x}/2$,
$L_{x}/2<x_{j}<L_{x}$, shown in Fig.\ref{fig2}(d). Therefore, all
ground states correspond to a condensate with {\it antiferromagnetic
order of quantum nature} (quantum nature means that the site-magnetization
away from the domain wall has only two possible values). The {\it
antiferromagnetic order} is a direct result of the existence of
the four degenerate separated minima ${\bf Q_{1,2,3,4}}$, which
themselves are a result of the anticommutation relation
$\{\epsilon_{k}\tau_{x},\Lambda_{k}\tau_{z}\}=0$. Therefore,
The key to realize the interesting {\it antiferromagnetic order}
is to realize the Rashba-like SOC, $\Lambda_{k}\tau_{z}$, which
needs a non-Abelian gauge field currently beyond the realization
ability of experiments. If SOC is as usual commutative with kinetic
term, the momentum shift of the minima can reach the values of
${\bf Q_{i}}$ only when $\alpha/t$ goes to infinite which is hard
to realize. For general finite $\alpha/t$, just like continuous
systems \cite{C. Wang, T.-L. Ho}, the lattice system will exhibit
site-dependent magnetization which is a classical quantity, but
can not establish the {\it antiferromagnetic order} of quantum
nature like here, which exists in a wide range of parameters.

{\it Discussions and Conclusions.---}
Duo to the existence of scattering processes related to the
reciprocal lattice vector, it is found that the ground states
always favor a coherent condensate and always  exhibit very large
degeneracy. However, this conclusion should only be valid when the
effect of quantum fluctuations is small and the phase keeps
coherent. When phase coherence is lost (then
the ground state energy is obtained by averaging the phases
\cite{E. J. Mueller}), the degeneracy will be greatly reduced and
the ground state is inclined to a {\it single} condensate. This
crossover can be observed by time-of-flight experiments.
As with fixed parameters, all ground states
correspond to the same spin configuration, which suggests that the
spin configurations are in fact more stable than the ground states.
The spin configurations can be revealed by spin-dependent imaging techniques.
The minima where the bosons are condensed and the shift of the minima
when $\alpha$ goes across $\alpha_{c}$ can also be observed by
time-of-flight experiments.

The coexistence of superfluidity and antiferromagnetic order in
a cold atomic system, which are the most two important phases in
high-$T_{c}$ superconductors, opens a door to study their interplay
in a controllable way.

{\it Acknowledgments.---} This work was supported by NSFC Grant No.11275180.

\begin{widetext}

\section{Supplementary Materials}
\subsection{A. Interaction under new representation.}
For Hamiltonian (\ref{1}), by redefining a representation $\tilde{\Phi}_{k}=(\hat{\alpha}_{1k}, \hat{\alpha}_{2k},
\hat{\beta}_{1k}, \hat{\beta}_{2k})^{T}=U(k)\Phi_{k}$, where $U(k)$ is a $4\times4$ matrix with the form
\begin{eqnarray}
U(k)=\left(\begin{array}{cccc}
             \frac{E(k)+\Delta\mu}{\mathcal{N}_{1}} & 0 & \frac{-E(k)+\Delta\mu}{\mathcal{N}_{2}} & 0 \\
             -\frac{A(k)}{\mathcal{N}_{1}} & -\frac{\epsilon_{k}}{\mathcal{N}_{1}} & -\frac{A(k)}{\mathcal{N}_{2}} & -\frac{\epsilon_{k}}{\mathcal{N}_{2}} \\
             -\frac{\epsilon_{k}}{\mathcal{N}_{1}} & \frac{A^{*}(k)}{\mathcal{N}_{1}} & -\frac{\epsilon_{k}}{\mathcal{N}_{2}} & \frac{A^{*}(k)}{\mathcal{N}_{2}}\\
             0 & \frac{E(k)+\Delta\mu}{\mathcal{N}_{1}} & 0 & \frac{-E(k)+\Delta\mu}{\mathcal{N}_{2}}
           \end{array}\right)^{-1}
\end{eqnarray}
where $E(k)=\sqrt{(\Delta\mu)^{2}+\epsilon_{k}^{2}+\alpha^{2}(\sin^{2}(k_{x}a)+\sin^{2}(k_{y}a))}$,
$A(k)=i\alpha\sin(k_{x}a)-\alpha\sin(k_{y}a)$, $\epsilon_{k}=-t(\cos(k_{x}a)+\cos(k_{y}a))$,
$\mathcal{N}_{1}=\sqrt{2E(k)(E(k)+\Delta\mu)}$, $\mathcal{N}_{2}=\sqrt{2E(k)(E(k)-\Delta\mu)}$,
then the Hamiltonian is diagonalized as
\begin{eqnarray}
H_{0}=\sum_{k}[-E(k)(\hat{\alpha}_{1k}^{\dag}\hat{\alpha}_{1k}+\hat{\alpha}_{2k}^{\dag}\hat{\alpha}_{2k})
+E(k)(\hat{\beta}_{1k}^{\dag}\hat{\beta}_{1k}+\hat{\beta}_{2k}^{\dag}\hat{\beta}_{2k})].
\end{eqnarray}

In the following, we set $\lambda_{1}(k)=\frac{E(k)+\Delta\mu}{\mathcal{N}_{1}}$, $\lambda_{2}(k)=\frac{-E(k)+\Delta\mu}{\mathcal{N}_{2}}$,
$\lambda_{3}=\frac{A(k)}{\mathcal{N}_{1}}$, $\lambda_{4}=\frac{\epsilon_{k}}{\mathcal{N}_{1}}$,
$\lambda_{5}=\frac{A(k)}{\mathcal{N}_{2}}$, $\lambda_{6}=\frac{\epsilon_{k}}{\mathcal{N}_{2}}$.
Then $\hat{a}_{k\uparrow}=\lambda_{1}(k)\hat{\alpha}_{1k}+\lambda_{2}(k)\hat{\beta}_{1k}$,
$\hat{a}_{k\downarrow}=\lambda_{1}(k)\hat{\alpha}_{2k}+\lambda_{2}(k)\hat{\beta}_{2k}$,
$\hat{b}_{k\downarrow}=-(\lambda_{3}(k)\hat{\alpha}_{1k}+\lambda_{4}(k)\hat{\alpha}_{2k}+
\lambda_{5}(k)\hat{\beta}_{1k}+\lambda_{6}(k)\hat{\beta}_{2k})$,
$\hat{b}_{k\uparrow}=-(\lambda_{4}(k)\hat{\alpha}_{1k}-\lambda_{3}^{*}(k)\hat{\alpha}_{2k}+
\lambda_{6}(k)\hat{\beta}_{1k}-\lambda_{5}^{*}(k)\hat{\beta}_{2k})$.

Under this representation, the form of the interaction will turn out to be very complicated. The concrete
interaction forms for sublattices $A$ are
\begin{eqnarray}
&&g_{1A}\sum_{k_{1},k_{2},k_{3},k_{4}}\hat{a}^{\dag}_{k_{1}\uparrow}\hat{a}^{\dag}_{k_{2}\uparrow}\hat{a}_{k_{3}\uparrow}\hat{a}_{k_{4}\uparrow}\nonumber\\
&=&g_{1A}\sum_{k_{1},k_{2},k_{3},k_{4}}(\lambda_{1}^{*}(k_{1})\hat{\alpha}_{1k_{1}}^{\dag}+\lambda_{2}^{*}(k_{1})\hat{\beta}_{1k_{1}}^{\dag})
(\lambda_{1}^{*}(k_{2})\hat{\alpha}_{1k_{2}}^{\dag}+\lambda_{2}^{*}(k_{2})\hat{\beta}_{1k_{2}}^{\dag})
(\lambda_{1}(k_{3})\hat{\alpha}_{1k_{3}}+\lambda_{2}(k_{3})\hat{\beta}_{1k_{3}})
(\lambda_{1}(k_{4})\hat{\alpha}_{1k_{4}}+\lambda_{2}(k_{4})\hat{\beta}_{1k_{4}})\nonumber\\
&=&g_{1A}\sum_{k_{1},k_{2},k_{3},k_{4}}[\lambda_{1}^{*}(k_{1})\lambda_{1}^{*}(k_{2})\lambda_{1}(k_{3})
\lambda_{1}(k_{4})\hat{\alpha}_{1k_{1}}^{\dag}\hat{\alpha}_{1k_{2}}^{\dag}\hat{\alpha}_{1k_{3}}\hat{\alpha}_{1k_{4}}
+2\lambda_{1}^{*}(k_{1})\lambda_{1}^{*}(k_{2})\lambda_{1}(k_{3})\lambda_{2}(k_{4})\hat{\alpha}_{1k_{1}}^{\dag}
\hat{\alpha}_{1k_{2}}^{\dag}\hat{\alpha}_{1k_{3}}\hat{\beta}_{1k_{4}}\nonumber\\
&&\qquad\qquad\qquad+\lambda_{1}^{*}(k_{1})\lambda_{1}^{*}(k_{2})\lambda_{2}(k_{3})\lambda_{2}(k_{4})\hat{\alpha}_{1k_{1}}^{\dag}
\hat{\alpha}_{1k_{2}}^{\dag}\hat{\beta}_{1k_{3}}\hat{\beta}_{1k_{4}}
+2\lambda_{1}^{*}(k_{1})\lambda_{2}^{*}(k_{2})\lambda_{1}(k_{3})\lambda_{1}(k_{4})\hat{\alpha}_{1k_{1}}^{\dag}
\hat{\beta}_{1k_{2}}^{\dag}\hat{\alpha}_{1k_{3}}\hat{\alpha}_{1k_{4}}\nonumber\\
&&\qquad\qquad\qquad+4\lambda_{1}^{*}(k_{1})\lambda_{2}^{*}(k_{2})\lambda_{1}(k_{3})\lambda_{2}(k_{4})\hat{\alpha}_{1k_{1}}^{\dag}
\hat{\beta}_{1k_{2}}^{\dag}\hat{\alpha}_{1k_{3}}\hat{\beta}_{1k_{4}}
+2\lambda_{1}^{*}(k_{1})\lambda_{2}^{*}(k_{2})\lambda_{2}(k_{3})\lambda_{2}(k_{4})\hat{\alpha}_{1k_{1}}^{\dag}
\hat{\beta}_{1k_{2}}^{\dag}\hat{\beta}_{1k_{3}}\hat{\beta}_{1k_{4}}\nonumber\\
&&\qquad\qquad\qquad+\lambda_{2}^{*}(k_{1})\lambda_{2}^{*}(k_{2})\lambda_{1}(k_{3})\lambda_{1}(k_{4})\hat{\beta}_{1k_{1}}^{\dag}
\hat{\beta}_{1k_{2}}^{\dag}\hat{\alpha}_{1k_{3}}\hat{\alpha}_{1k_{4}}
+2\lambda_{2}^{*}(k_{1})\lambda_{2}^{*}(k_{2})\lambda_{1}(k_{3})\lambda_{2}(k_{4})\hat{\beta}_{1k_{1}}^{\dag}
\hat{\beta}_{1k_{2}}^{\dag}\hat{\alpha}_{1k_{3}}\hat{\beta}_{1k_{4}}\nonumber\\
&&\qquad\qquad\qquad+\lambda_{2}^{*}(k_{1})\lambda_{2}^{*}(k_{2})\lambda_{2}(k_{3})\lambda_{2}(k_{4})\hat{\beta}_{1k_{1}}^{\dag}
\hat{\beta}_{1k_{2}}^{\dag}\hat{\beta}_{1k_{3}}\hat{\beta}_{1k_{4}}],\nonumber
\end{eqnarray}

\begin{eqnarray}
&&g_{1A}\sum_{k_{1},k_{2},k_{3},k_{4}}\hat{a}^{\dag}_{k_{1}\downarrow}\hat{a}^{\dag}_{k_{2}\downarrow}\hat{a}_{k_{3}\downarrow}\hat{a}_{k_{4}\downarrow}\nonumber\\
&=&g_{1A}\sum_{k_{1},k_{2},k_{3},k_{4}}(\lambda_{1}^{*}(k_{1})\hat{\alpha}_{2k_{1}}^{\dag}+\lambda_{2}^{*}(k_{1})\hat{\beta}_{2k_{1}}^{\dag})
(\lambda_{1}^{*}(k_{2})\hat{\alpha}_{2k_{2}}^{\dag}+\lambda_{2}^{*}(k_{2})\hat{\beta}_{2k_{2}}^{\dag})
(\lambda_{1}(k_{3})\hat{\alpha}_{2k_{3}}+\lambda_{2}(k_{3})\hat{\beta}_{2k_{3}})
(\lambda_{1}(k_{4})\hat{\alpha}_{2k_{4}}+\lambda_{2}(k_{4})\hat{\beta}_{2k_{4}})\nonumber\\
&=&g_{1A}\sum_{k_{1},k_{2},k_{3},k_{4}}[\lambda_{1}^{*}(k_{1})\lambda_{1}^{*}(k_{2})\lambda_{1}(k_{3})
\lambda_{1}(k_{4})\hat{\alpha}_{2k_{1}}^{\dag}\hat{\alpha}_{2k_{2}}^{\dag}\hat{\alpha}_{2k_{3}}\hat{\alpha}_{2k_{4}}
+2\lambda_{1}^{*}(k_{1})\lambda_{1}^{*}(k_{2})\lambda_{1}(k_{3})\lambda_{2}(k_{4})\hat{\alpha}_{2k_{1}}^{\dag}
\hat{\alpha}_{2k_{2}}^{\dag}\hat{\alpha}_{2k_{3}}\hat{\beta}_{2k_{4}}\nonumber\\
&&\qquad\qquad\qquad+\lambda_{1}^{*}(k_{1})\lambda_{1}^{*}(k_{2})\lambda_{2}(k_{3})\lambda_{2}(k_{4})\hat{\alpha}_{2k_{1}}^{\dag}
\hat{\alpha}_{2k_{2}}^{\dag}\hat{\beta}_{2k_{3}}\hat{\beta}_{2k_{4}}
+2\lambda_{1}^{*}(k_{1})\lambda_{2}^{*}(k_{2})\lambda_{1}(k_{3})\lambda_{1}(k_{4})\hat{\alpha}_{2k_{1}}^{\dag}
\hat{\beta}_{2k_{2}}^{\dag}\hat{\alpha}_{2k_{3}}\hat{\alpha}_{2k_{4}}\nonumber\\
&&\qquad\qquad\qquad+4\lambda_{1}^{*}(k_{1})\lambda_{2}^{*}(k_{2})\lambda_{1}(k_{3})\lambda_{2}(k_{4})\hat{\alpha}_{2k_{1}}^{\dag}
\hat{\beta}_{2k_{2}}^{\dag}\hat{\alpha}_{2k_{3}}\hat{\beta}_{2k_{4}}
+2\lambda_{1}^{*}(k_{1})\lambda_{2}^{*}(k_{2})\lambda_{2}(k_{3})\lambda_{2}(k_{4})\hat{\alpha}_{2k_{1}}^{\dag}
\hat{\beta}_{2k_{2}}^{\dag}\hat{\beta}_{2k_{3}}\hat{\beta}_{2k_{4}}\nonumber\\
&&\qquad\qquad\qquad+\lambda_{2}^{*}(k_{1})\lambda_{2}^{*}(k_{2})\lambda_{1}(k_{3})\lambda_{1}(k_{4})\hat{\beta}_{2k_{1}}^{\dag}
\hat{\beta}_{2k_{2}}^{\dag}\hat{\alpha}_{2k_{3}}\hat{\alpha}_{2k_{4}}
+2\lambda_{2}^{*}(k_{1})\lambda_{2}^{*}(k_{2})\lambda_{1}(k_{3})\lambda_{2}(k_{4})\hat{\beta}_{2k_{1}}^{\dag}
\hat{\beta}_{2k_{2}}^{\dag}\hat{\alpha}_{2k_{3}}\hat{\beta}_{2k_{4}}\nonumber\\
&&\qquad\qquad\qquad+\lambda_{2}^{*}(k_{1})\lambda_{2}^{*}(k_{2})\lambda_{2}(k_{3})\lambda_{2}(k_{4})\hat{\beta}_{2k_{1}}^{\dag}
\hat{\beta}_{2k_{2}}^{\dag}\hat{\beta}_{2k_{3}}\hat{\beta}_{2k_{4}}],\nonumber
\end{eqnarray}

\begin{eqnarray}
&&2g_{12A}\sum_{k_{1},k_{2},k_{3},k_{4}}\hat{a}^{\dag}_{k_{1}\downarrow}\hat{a}^{\dag}_{k_{2}\downarrow}\hat{a}_{k_{3}\downarrow}\hat{a}_{k_{4}\downarrow}\nonumber\\
&=&g_{12A}\sum_{k_{1},k_{2},k_{3},k_{4}}(\lambda_{1}^{*}(k_{1})\hat{\alpha}_{1k_{1}}^{\dag}+\lambda_{2}^{*}(k_{1})\hat{\beta}_{1k_{1}}^{\dag})
(\lambda_{1}^{*}(k_{2})\hat{\alpha}_{2k_{2}}^{\dag}+\lambda_{2}^{*}(k_{2})\hat{\beta}_{2k_{2}}^{\dag})
(\lambda_{1}(k_{3})\hat{\alpha}_{2k_{3}}+\lambda_{2}(k_{3})\hat{\beta}_{2k_{3}})
(\lambda_{1}(k_{4})\hat{\alpha}_{1k_{4}}+\lambda_{2}(k_{4})\hat{\beta}_{1k_{4}})\nonumber\\
&=&2g_{12A}\sum_{k_{1},k_{2},k_{3},k_{4}}[\lambda_{1}^{*}(k_{1})\lambda_{1}^{*}(k_{2})\lambda_{1}(k_{3})
\lambda_{1}(k_{4})\hat{\alpha}_{1k_{1}}^{\dag}\hat{\alpha}_{2k_{2}}^{\dag}\hat{\alpha}_{2k_{3}}\hat{\alpha}_{1k_{4}}
+\lambda_{1}^{*}(k_{1})\lambda_{1}^{*}(k_{2})\lambda_{1}(k_{3})\lambda_{2}(k_{4})\hat{\alpha}_{1k_{1}}^{\dag}
\hat{\alpha}_{2k_{2}}^{\dag}(\hat{\alpha}_{2k_{3}}\hat{\beta}_{1k_{4}}+\hat{\alpha}_{1k_{3}}\hat{\beta}_{2k_{4}})\nonumber\\
&&\qquad\qquad\qquad+\lambda_{1}^{*}(k_{1})\lambda_{1}^{*}(k_{2})\lambda_{2}(k_{3})\lambda_{2}(k_{4})\hat{\alpha}_{1k_{1}}^{\dag}
\hat{\alpha}_{2k_{2}}^{\dag}\hat{\beta}_{2k_{3}}\hat{\beta}_{1k_{4}}
+\lambda_{1}^{*}(k_{1})\lambda_{2}^{*}(k_{2})\lambda_{1}(k_{3})\lambda_{1}(k_{4})(\hat{\alpha}_{1k_{1}}^{\dag}
\hat{\beta}_{2k_{2}}^{\dag}+\hat{\alpha}_{2k_{1}}^{\dag}\hat{\beta}_{1k_{2}}^{\dag})\hat{\alpha}_{2k_{3}}\hat{\alpha}_{1k_{4}}\nonumber\\
&&\qquad\qquad\qquad+\lambda_{1}^{*}(k_{1})\lambda_{2}^{*}(k_{2})\lambda_{1}(k_{3})\lambda_{2}(k_{4})(\hat{\alpha}_{1k_{1}}^{\dag}
\hat{\beta}_{2k_{2}}^{\dag}+\hat{\alpha}_{2k_{1}}^{\dag}\hat{\beta}_{1k_{2}}^{\dag})
(\hat{\alpha}_{2k_{3}}\hat{\beta}_{1k_{4}}+\hat{\alpha}_{1k_{3}}\hat{\beta}_{2k_{4}})
+\lambda_{1}^{*}(k_{1})\lambda_{2}^{*}(k_{2})\lambda_{2}(k_{3})\lambda_{2}(k_{4})\nonumber\\
&&\qquad\qquad\qquad(\hat{\alpha}_{1k_{1}}^{\dag}
\hat{\beta}_{2k_{2}}^{\dag}+\hat{\alpha}_{2k_{1}}^{\dag}\hat{\beta}_{1k_{2}}^{\dag})\hat{\beta}_{2k_{3}}\hat{\beta}_{2k_{4}}
+\lambda_{2}^{*}(k_{1})\lambda_{2}^{*}(k_{2})\lambda_{1}(k_{3})\lambda_{1}(k_{4})\hat{\beta}_{1k_{1}}^{\dag}
\hat{\beta}_{2k_{2}}^{\dag}\hat{\alpha}_{2k_{3}}\hat{\alpha}_{1k_{4}}
\nonumber\\
&&\qquad\qquad\qquad
+\lambda_{2}^{*}(k_{1})\lambda_{2}^{*}(k_{2})\lambda_{1}(k_{3})\lambda_{2}(k_{4})\hat{\beta}_{1k_{1}}^{\dag}
\hat{\beta}_{2k_{2}}^{\dag}(\hat{\alpha}_{1k_{1}}^{\dag}\hat{\beta}_{2k_{2}}^{\dag}+\hat{\alpha}_{2k_{1}}^{\dag}\hat{\beta}_{1k_{2}}^{\dag})
+\lambda_{2}^{*}(k_{1})\lambda_{2}^{*}(k_{2})\lambda_{2}(k_{3})\lambda_{2}(k_{4})\hat{\beta}_{1k_{1}}^{\dag}
\hat{\beta}_{2k_{2}}^{\dag}\hat{\beta}_{2k_{3}}\hat{\beta}_{1k_{4}}],\nonumber
\end{eqnarray}
where the four momentums ${\bf k}_{1,2,3,4}$ for summation belong to the first Brillouin zone
and need to satisfy the constraint ${\bf k}_{1}+{\bf k}_{2}-{\bf k}_{3}-{\bf k}_{4}=n{\bf G}$
with $n$ an integer.

As the expressions of $\hat{b}_{k\uparrow,\downarrow}$ have even more terms, under the new
representation, the expressions for interactions on the sublattices $B$ will turns out to
be too complicated. As the higher band almost has no effect on the ground state when the
temperature is low, in fact we can neglect terms involving $\hat{\beta}_{1,2}$. Although
in continuous system, it is found that the higher band can induce divergent effective interaction
in the lower band \cite{T. Ozawa1}, here the divergence behavior will be avoided since there is a natural
cutoff, $2\pi/a$, for momentum, and if we consider the third dimension which is strongly confined,
the infrared divergence is also absent. Therefore, in the following, for simplicity, we neglect all
terms involving $\hat{\beta}_{1,2}$ and do not consider the renormalization of the
interaction.

\begin{eqnarray}
&&g_{1B}\sum_{k_{1},k_{2},k_{3},k_{4}}\hat{b}^{\dag}_{k_{1}\uparrow}\hat{b}^{\dag}_{k_{2}\uparrow}\hat{b}_{k_{3}\uparrow}\hat{b}_{k_{4}\uparrow}\nonumber\\
&=&g_{1B}\sum_{k_{1},k_{2},k_{3},k_{4}}(\lambda_{4}^{*}(k_{1})\hat{\alpha}_{1k_{1}}^{\dag}-\lambda_{3}(k_{1})\hat{\alpha}_{2k_{1}}^{\dag})
(\lambda_{4}^{*}(k_{2})\hat{\alpha}_{1k_{2}}^{\dag}-\lambda_{3}(k_{2})\hat{\alpha}_{2k_{2}}^{\dag})
(\lambda_{4}(k_{3})\hat{\alpha}_{1k_{3}}-\lambda_{3}^{*}(k_{3})\hat{\alpha}_{2k_{3}})
(\lambda_{4}(k_{4})\hat{\alpha}_{1k_{4}}-\lambda_{3}^{*}(k_{4})\hat{\alpha}_{2k_{4}})\nonumber\\
&=&g_{1B}\sum_{k_{1},k_{2},k_{3},k_{4}}[\lambda_{4}^{*}(k_{1})\lambda_{4}^{*}(k_{2})\lambda_{4}(k_{3})
\lambda_{4}(k_{4})\hat{\alpha}_{1k_{1}}^{\dag}\hat{\alpha}_{1k_{2}}^{\dag}\hat{\alpha}_{1k_{3}}\hat{\alpha}_{1k_{4}}
-2\lambda_{4}^{*}(k_{1})\lambda_{4}^{*}(k_{2})\lambda_{4}(k_{3})\lambda_{3}^{*}(k_{4})\hat{\alpha}_{1k_{1}}^{\dag}
\hat{\alpha}_{1k_{2}}^{\dag}\hat{\alpha}_{1k_{3}}\hat{\alpha}_{2k_{4}}\nonumber\\
&&\qquad\qquad\qquad+\lambda_{4}^{*}(k_{1})\lambda_{4}^{*}(k_{2})\lambda_{3}(k_{3})^{*}\lambda_{3}^{*}(k_{4})\hat{\alpha}_{1k_{1}}^{\dag}
\hat{\alpha}_{1k_{2}}^{\dag}\hat{\alpha}_{2k_{3}}\hat{\alpha}_{2k_{4}}
-2\lambda_{4}^{*}(k_{1})\lambda_{3}(k_{2})\lambda_{4}(k_{3})\lambda_{4}(k_{4})\hat{\alpha}_{1k_{1}}^{\dag}
\hat{\alpha}_{2k_{2}}^{\dag}\hat{\alpha}_{1k_{3}}\hat{\alpha}_{1k_{4}}\nonumber\\
&&\qquad\qquad\qquad+4\lambda_{4}^{*}(k_{1})\lambda_{3}(k_{2})\lambda_{4}(k_{3})\lambda_{3}^{*}(k_{4})\hat{\alpha}_{1k_{1}}^{\dag}
\hat{\alpha}_{2k_{2}}^{\dag}\hat{\alpha}_{1k_{3}}\hat{\alpha}_{2k_{4}}
-2\lambda_{4}^{*}(k_{1})\lambda_{3}(k_{2})\lambda_{3}^{*}(k_{3})\lambda_{3}^{*}(k_{4})\hat{\alpha}_{1k_{1}}^{\dag}
\hat{\alpha}_{2k_{2}}^{\dag}\hat{\alpha}_{2k_{3}}\hat{\alpha}_{2k_{4}}\nonumber\\
&&\qquad\qquad\qquad+\lambda_{3}(k_{1})\lambda_{3}(k_{2})\lambda_{4}(k_{3})\lambda_{4}(k_{4})\hat{\alpha}_{2k_{1}}^{\dag}
\hat{\alpha}_{2k_{2}}^{\dag}\hat{\alpha}_{1k_{3}}\hat{\alpha}_{1k_{4}}
-2\lambda_{3}(k_{1})\lambda_{3}(k_{2})\lambda_{4}(k_{3})\lambda_{3}^{*}(k_{4})\hat{\alpha}_{2k_{1}}^{\dag}
\hat{\alpha}_{2k_{2}}^{\dag}\hat{\alpha}_{1k_{3}}\hat{\alpha}_{2k_{4}}\nonumber\\
&&\qquad\qquad\qquad+\lambda_{3}(k_{1})\lambda_{3}(k_{2})\lambda_{3}^{*}(k_{3})\lambda_{3}^{*}(k_{4})\hat{\alpha}_{2k_{1}}^{\dag}
\hat{\alpha}_{2k_{2}}^{\dag}\hat{\alpha}_{2k_{3}}\hat{\alpha}_{2k_{4}}],\nonumber
\end{eqnarray}

\begin{eqnarray}
&&g_{1B}\sum_{k_{1},k_{2},k_{3},k_{4}}\hat{b}^{\dag}_{k_{1}\downarrow}\hat{b}^{\dag}_{k_{2}\downarrow}\hat{b}_{k_{3}\downarrow}\hat{b}_{k_{4}\downarrow}\nonumber\\
&=&g_{1B}\sum_{k_{1},k_{2},k_{3},k_{4}}(\lambda_{3}^{*}(k_{1})\hat{\alpha}_{1k_{1}}^{\dag}+\lambda_{4}^{*}(k_{1})\hat{\alpha}_{2k_{1}}^{\dag})
(\lambda_{3}^{*}(k_{2})\hat{\alpha}_{1k_{2}}^{\dag}+\lambda_{4}^{*}(k_{2})\hat{\alpha}_{2k_{2}}^{\dag})
(\lambda_{3}(k_{4})\hat{\alpha}_{1k_{4}}+\lambda_{4}(k_{4})\hat{\alpha}_{2k_{4}})
(\lambda_{3}(k_{3})\hat{\alpha}_{1k_{3}}+\lambda_{4}(k_{3})\hat{\alpha}_{2k_{3}})\nonumber\\
&=&g_{1B}\sum_{k_{1},k_{2},k_{3},k_{4}}[\lambda_{3}^{*}(k_{1})\lambda_{3}^{*}(k_{2})\lambda_{3}(k_{3})
\lambda_{3}(k_{4})\hat{\alpha}_{1k_{1}}^{\dag}\hat{\alpha}_{1k_{2}}^{\dag}\hat{\alpha}_{1k_{3}}\hat{\alpha}_{1k_{4}}
+2\lambda_{3}^{*}(k_{1})\lambda_{3}^{*}(k_{2})\lambda_{3}(k_{3})\lambda_{4}(k_{4})\hat{\alpha}_{1k_{1}}^{\dag}
\hat{\alpha}_{1k_{2}}^{\dag}\hat{\alpha}_{1k_{3}}\hat{\alpha}_{2k_{4}}\nonumber\\
&&\qquad\qquad\qquad+\lambda_{3}^{*}(k_{1})\lambda_{3}^{*}(k_{2})\lambda_{4}(k_{3})\lambda_{4}(k_{4})\hat{\alpha}_{1k_{1}}^{\dag}
\hat{\alpha}_{1k_{2}}^{\dag}\hat{\alpha}_{2k_{3}}\hat{\alpha}_{2k_{4}}
+2\lambda_{3}^{*}(k_{1})\lambda_{4}^{*}(k_{2})\lambda_{3}(k_{3})\lambda_{3}(k_{4})\hat{\alpha}_{1k_{1}}^{\dag}
\hat{\alpha}_{2k_{2}}^{\dag}\hat{\alpha}_{1k_{3}}\hat{\alpha}_{1k_{4}}\nonumber\\
&&\qquad\qquad\qquad+4\lambda_{3}^{*}(k_{1})\lambda_{4}^{*}(k_{2})\lambda_{3}(k_{3})\lambda_{4}(k_{4})\hat{\alpha}_{1k_{1}}^{\dag}
\hat{\alpha}_{2k_{2}}^{\dag}\hat{\alpha}_{1k_{3}}\hat{\alpha}_{2k_{4}}
+2\lambda_{3}^{*}(k_{1})\lambda_{4}^{*}(k_{2})\lambda_{4}(k_{3})\lambda_{4}(k_{4})\hat{\alpha}_{1k_{1}}^{\dag}
\hat{\alpha}_{2k_{2}}^{\dag}\hat{\alpha}_{2k_{3}}\hat{\alpha}_{2k_{4}}\nonumber\\
&&\qquad\qquad\qquad+\lambda_{4}^{*}(k_{1})\lambda_{4}^{*}(k_{2})\lambda_{3}(k_{3})\lambda_{3}(k_{4})\hat{\alpha}_{2k_{1}}^{\dag}
\hat{\alpha}_{2k_{2}}^{\dag}\hat{\alpha}_{1k_{3}}\hat{\alpha}_{1k_{4}}
+2\lambda_{4}^{*}(k_{1})\lambda_{4}^{*}(k_{2})\lambda_{3}(k_{3})\lambda_{4}(k_{4})\hat{\alpha}_{2k_{1}}^{\dag}
\hat{\alpha}_{2k_{2}}^{\dag}\hat{\alpha}_{1k_{3}}\hat{\alpha}_{2k_{4}}\nonumber\\
&&\qquad\qquad\qquad+\lambda_{4}^{*}(k_{1})\lambda_{4}^{*}(k_{2})\lambda_{4}(k_{3})\lambda_{4}(k_{4})\hat{\alpha}_{2k_{1}}^{\dag}
\hat{\alpha}_{2k_{2}}^{\dag}\hat{\alpha}_{2k_{3}}\hat{\alpha}_{2k_{4}}],\nonumber
\end{eqnarray}

\begin{eqnarray}
&&2g_{12B}\sum_{k_{1},k_{2},k_{3},k_{4}}\hat{b}^{\dag}_{k_{1}\uparrow}\hat{b}^{\dag}_{k_{2}\downarrow}\hat{b}_{k_{3}\downarrow}\hat{b}_{k_{4}\uparrow}\nonumber\\
&=&2g_{12B}\sum_{k_{1},k_{2},k_{3},k_{4}}(\lambda_{4}^{*}(k_{1})\hat{\alpha}_{1k_{1}}^{\dag}-\lambda_{3}(k_{1})\hat{\alpha}_{2k_{1}}^{\dag})
(\lambda_{3}^{*}(k_{2})\hat{\alpha}_{1k_{2}}^{\dag}+\lambda_{4}^{*}(k_{2})\hat{\alpha}_{2k_{2}}^{\dag})
(\lambda_{3}(k_{3})\hat{\alpha}_{1k_{3}}+\lambda_{4}(k_{3})\hat{\alpha}_{2k_{3}})
(\lambda_{4}(k_{4})\hat{\alpha}_{1k_{4}}-\lambda_{3}^{*}(k_{4})\hat{\alpha}_{2k_{4}})\nonumber\\
&=&2g_{12B}\sum_{k_{1},k_{2},k_{3},k_{4}}[\lambda_{4}^{*}(k_{1})\lambda_{3}^{*}(k_{2})\lambda_{3}(k_{3})
\lambda_{4}(k_{4})\hat{\alpha}_{1k_{1}}^{\dag}\hat{\alpha}_{1k_{2}}^{\dag}\hat{\alpha}_{1k_{3}}\hat{\alpha}_{1k_{4}}
+\lambda_{4}^{*}(k_{1})\lambda_{3}^{*}(k_{2})(\lambda_{4}(k_{3})\lambda_{4}(k_{4})-\lambda_{3}(k_{3})\lambda_{3}^{*}(k_{4}))\nonumber\\
&&\qquad\qquad\qquad\times\hat{\alpha}_{1k_{1}}^{\dag}
\hat{\alpha}_{1k_{2}}^{\dag}\hat{\alpha}_{1k_{3}}\hat{\alpha}_{2k_{4}}
-\lambda_{4}^{*}(k_{1})\lambda_{3}^{*}(k_{2})\lambda_{4}(k_{3})\lambda_{3}^{*}(k_{4})\hat{\alpha}_{1k_{1}}^{\dag}
\hat{\alpha}_{1k_{2}}^{\dag}\hat{\alpha}_{2k_{3}}\hat{\alpha}_{2k_{4}}
+(\lambda_{4}^{*}(k_{1})\lambda_{4}^{*}(k_{2})-\lambda_{3}^{*}(k_{1})\lambda_{3}(k_{2}))
\nonumber\\
&&\qquad\qquad\qquad
\times\lambda_{3}(k_{3})\lambda_{4}(k_{4})\hat{\alpha}_{1k_{1}}^{\dag}
\hat{\alpha}_{2k_{2}}^{\dag}\hat{\alpha}_{1k_{3}}\hat{\alpha}_{1k_{4}}
+(\lambda_{4}^{*}(k_{1})\lambda_{4}^{*}(k_{2})-\lambda_{3}^{*}(k_{1})\lambda_{3}(k_{2}))
(\lambda_{4}(k_{3})\lambda_{4}(k_{4})-\lambda_{3}^{*}(k_{3})\lambda_{3}(k_{4}))\nonumber\\
&&\qquad\qquad\qquad\times\hat{\alpha}_{1k_{1}}^{\dag}
\hat{\alpha}_{2k_{2}}^{\dag}\hat{\alpha}_{2k_{3}}\hat{\alpha}_{1k_{4}}
-(\lambda_{4}^{*}(k_{1})\lambda_{4}^{*}(k_{2})-\lambda_{3}^{*}(k_{1})\lambda_{3}(k_{2}))\lambda_{3}(k_{3})\lambda_{3}^{*}(k_{4})\hat{\alpha}_{1k_{1}}^{\dag}
\hat{\alpha}_{2k_{2}}^{\dag}\hat{\alpha}_{2k_{3}}\hat{\alpha}_{2k_{4}}\nonumber\\
&&\qquad\qquad\qquad-\lambda_{3}(k_{1})\lambda_{4}^{*}(k_{2})\lambda_{3}(k_{3})\lambda_{4}(k_{4})\hat{\alpha}_{2k_{1}}^{\dag}
\hat{\alpha}_{2k_{2}}^{\dag}\hat{\alpha}_{1k_{3}}\hat{\alpha}_{1k_{4}}
-\lambda_{3}(k_{1})\lambda_{4}^{*}(k_{2})(\lambda_{4}(k_{3})\lambda_{4}(k_{4})-\lambda_{3}(k_{3})\lambda_{3}^{*}(k_{4}))\nonumber\\
&&\qquad\qquad\qquad\times\hat{\alpha}_{2k_{1}}^{\dag}
\hat{\alpha}_{2k_{2}}^{\dag}\hat{\alpha}_{1k_{1}}^{\dag}\hat{\alpha}_{2k_{2}}^{\dag}
+\lambda_{3}(k_{1})\lambda_{4}^{*}(k_{2})\lambda_{4}(k_{3})\lambda_{3}^{*}(k_{4})\hat{\alpha}_{2k_{1}}^{\dag}
\hat{\alpha}_{2k_{2}}^{\dag}\hat{\alpha}_{2k_{3}}\hat{\alpha}_{2k_{4}}].
\end{eqnarray}

Although the interaction forms are very complicated, we only need
to consider several of them when we are going to determine the
ground state. For example, when the bosons are condensed at ${\bf
k_{0}}$ or ${\bf k_{\pi}}$, as $\lambda_{3}({\bf
k_{0}})=\lambda_{3}({\bf k_{\pi}})=0$, in fact only the following
terms have contribution to the energy of ground state,
\begin{eqnarray}
&&\sum_{k_{1},k_{2},k_{3},k_{4}}\{\lambda_{1}(k_{1})\lambda_{1}(k_{2})\lambda_{1}(k_{3})
\lambda_{1}(k_{4})[g_{1A}(\hat{\alpha}_{1k_{1}}^{\dag}\hat{\alpha}_{1k_{2}}^{\dag}\hat{\alpha}_{1k_{3}}\hat{\alpha}_{1k_{4}}+
\hat{\alpha}_{2k_{1}}^{\dag}\hat{\alpha}_{2k_{2}}^{\dag}\hat{\alpha}_{2k_{3}}\hat{\alpha}_{2k_{4}})+
2g_{12A}\hat{\alpha}_{1k_{1}}^{\dag}
\hat{\alpha}_{2k_{2}}^{\dag}\hat{\alpha}_{2k_{3}}\hat{\alpha}_{1k_{4}}]\nonumber\\
&&\qquad\qquad+\lambda_{4}(k_{1})\lambda_{4}(k_{2})\lambda_{4}(k_{3})
\lambda_{4}(k_{4})[g_{1B}(\hat{\alpha}_{1k_{1}}^{\dag}\hat{\alpha}_{1k_{2}}^{\dag}\hat{\alpha}_{1k_{3}}\hat{\alpha}_{1k_{4}}+
\hat{\alpha}_{2k_{1}}^{\dag}\hat{\alpha}_{2k_{2}}^{\dag}\hat{\alpha}_{2k_{3}}\hat{\alpha}_{2k_{4}})+
2g_{12B}\hat{\alpha}_{1k_{1}}^{\dag}
\hat{\alpha}_{2k_{2}}^{\dag}\hat{\alpha}_{2k_{3}}\hat{\alpha}_{1k_{4}}]\}.
\end{eqnarray}
Therefore, the calculation is in fact not very tedious.

\subsection{B. Ground state energy for $\alpha>\alpha_{c}$.}
When $\alpha>\alpha_{c}$, the bosons will be condensed at ${\bf Q_{1,2,3,4}}$. As $\lambda_{4}({\bf Q_{1,2,3,4}})=0$,
the terms that have contribution to the ground state energy are given as
\begin{eqnarray}
&&\sum_{k_{1},k_{2},k_{3},k_{4}}\{\lambda_{1}(k_{1})\lambda_{1}(k_{2})\lambda_{1}(k_{3})
\lambda_{1}(k_{4})[g_{1A}(\hat{\alpha}_{1k_{1}}^{\dag}\hat{\alpha}_{1k_{2}}^{\dag}\hat{\alpha}_{1k_{3}}\hat{\alpha}_{1k_{4}}+
\hat{\alpha}_{2k_{1}}^{\dag}\hat{\alpha}_{2k_{2}}^{\dag}\hat{\alpha}_{2k_{3}}\hat{\alpha}_{2k_{4}})+
2g_{12A}\hat{\alpha}_{1k_{1}}^{\dag}
\hat{\alpha}_{2k_{2}}^{\dag}\hat{\alpha}_{2k_{3}}\hat{\alpha}_{1k_{4}}]\nonumber\\
&&\qquad\qquad+g_{1B}(\lambda_{3}^{*}(k_{1})\lambda_{3}^{*}(k_{2})\lambda_{3}(k_{3})
\lambda_{3}(k_{4})\hat{\alpha}_{1k_{1}}^{\dag}\hat{\alpha}_{1k_{2}}^{\dag}\hat{\alpha}_{1k_{3}}\hat{\alpha}_{1k_{4}}+
\lambda_{3}(k_{1})\lambda_{3}(k_{2})\lambda_{3}^{*}(k_{3})\lambda_{3}^{*}(k_{4})\hat{\alpha}_{2k_{1}}^{\dag}
\hat{\alpha}_{2k_{2}}^{\dag}\hat{\alpha}_{2k_{3}}\hat{\alpha}_{2k_{4}})\nonumber\\
&&\qquad\qquad+2g_{12B}(\lambda_{3}^{*}(k_{1})\lambda_{3}(k_{2})\lambda_{3}^{*}(k_{3})\lambda_{3}(k_{4})
\hat{\alpha}_{1k_{1}}^{\dag}\hat{\alpha}_{2k_{2}}^{\dag}\hat{\alpha}_{2k_{3}}\hat{\alpha}_{1k_{4}})\}.\label{22}
\end{eqnarray}

Combining Eq.(\ref{13}) and Eq.(\ref{22}), it is direct to obtain $<\tilde{\Psi}_{c}|H_{int}|\tilde{\Psi}_{c}>$
whose concrete form is given as
\begin{eqnarray}
<\tilde{\Psi}_{c}|H_{int}|\tilde{\Psi}_{c}>&=&<\tilde{\Psi}_{f}|H_{int}|\tilde{\Psi}_{f}>+\{2U_{1,A}\frac{N_{1}-1}{N_{1}}
[N_{11}N_{12}\cos(2\phi_{1,2})+N_{11}N_{13}\cos(2\phi_{1,3})+N_{11}N_{14}\cos(2\phi_{1,4})\nonumber\\
&&+N_{12}N_{13}\cos2(\phi_{1,3}-\phi_{1,2})+N_{12}N_{14}\cos2(\phi_{1,4}-\phi_{1,2})+N_{13}N_{14}\cos2(\phi_{1,4}-\phi_{1,3})\nonumber\\
&&+4\sqrt{N_{11}N_{12}N_{13}N_{14}}[\cos(\phi_{1,4}+\phi_{1,3}-\phi_{1,2})+\cos(\phi_{1,2}+\phi_{1,3}-\phi_{1,4})+\cos(\phi_{1,2}+\phi_{1,4}-\phi_{1,3})]]\nonumber\\
&&+2U_{1,A}\frac{N_{2}-1}{N_{2}}[N_{21}N_{22}\cos(2\phi_{2,2})+N_{21}N_{23}\cos(2\phi_{2,3})
+N_{21}N_{24}\cos(2\phi_{2,4})+N_{22}N_{23}\cos2(\phi_{2,3}-\phi_{2,2})\nonumber\\
&&+N_{22}N_{24}\cos2(\phi_{2,4}-\phi_{2,2})+N_{23}N_{24}\cos2(\phi_{2,4}-\phi_{2,3})+4\sqrt{N_{21}N_{22}N_{23}N_{24}}[\cos(\phi_{2,4}+\phi_{2,3}-\phi_{2,2})\nonumber\\
&&+\cos(\phi_{2,2}+\phi_{2,3}-\phi_{2,4})+\cos(\phi_{2,2}+\phi_{2,4}-\phi_{2,3})]]+8U_{2,A}[\sqrt{N_{11}N_{12}N_{21}N_{22}}\cos(\phi_{1,2})\cos(\phi_{2,2})\nonumber\\
&&+\sqrt{N_{11}N_{13}N_{21}N_{23}}\cos(\phi_{1,3})\cos(\phi_{2,3})+\sqrt{N_{11}N_{14}N_{21}N_{24}}\cos(\phi_{1,4})\cos(\phi_{2,4})+
\sqrt{N_{12}N_{13}N_{22}N_{23}}\nonumber\\
&&*\cos(\phi_{1,3}-\phi_{1,2})\cos(\phi_{2,3}-\phi_{2,2})+\sqrt{N_{12}N_{14}N_{22}N_{24}}\cos(\phi_{1,4}-\phi_{1,2})\cos(\phi_{2,4}-\phi_{2,2})\nonumber\\
&&+\sqrt{N_{13}N_{14}N_{22}N_{23}}\cos(\phi_{1,4}-\phi_{1,3})\cos(\phi_{2,4}-\phi_{2,3})]+8U_{2,A}
[\sqrt{N_{11}N_{12}N_{23}N_{24}}\cos(\phi_{1,2})\cos(\phi_{2,4}-\phi_{2,3})\nonumber\\
&&+\sqrt{N_{11}N_{13}N_{22}N_{24}}\cos(\phi_{1,3})\cos(\phi_{2,4}-\phi_{2,2})+\sqrt{N_{11}N_{14}N_{22}N_{23}}\cos(\phi_{1,4})\cos(\phi_{2,3}-\phi_{2,2})\nonumber\\
&&+\sqrt{N_{12}N_{13}N_{21}N_{24}}\cos(\phi_{1,3}-\phi_{1,2})\cos(\phi_{2,4})+\sqrt{N_{12}N_{14}N_{21}N_{23}}\cos(\phi_{1,4}-\phi_{1,2})\cos(\phi_{2,3})\nonumber\\
&&+\sqrt{N_{13}N_{14}N_{21}N_{22}}\cos(\phi_{1,4}-\phi_{1,3})\cos(\phi_{2,2})]\}\cos^{4}(\tilde{\theta}/2)+\{2U_{1,B}\frac{N_{1}-1}{N_{1}}[
N_{11}N_{12}\cos(2\phi_{1,2})\nonumber\\
&&-N_{11}N_{13}\cos(2\phi_{1,3})-N_{11}N_{14}\cos(2\phi_{1,4})-N_{12}N_{13}\cos2(\phi_{1,3}-\phi_{1,2})-N_{12}N_{14}\cos2(\phi_{1,4}-\phi_{1,2})\nonumber\\
&&+N_{13}N_{14}\cos2(\phi_{1,4}-\phi_{1,3})+4\sqrt{N_{11}N_{12}N_{13}N_{14}}[-\cos(\phi_{1,4}+\phi_{1,3}-\phi_{1,2})+\cos(\phi_{1,2}+\phi_{1,3}-\phi_{1,4})\nonumber\\
&&+\cos(\phi_{1,2}+\phi_{1,4}-\phi_{1,3})]]+2U_{1,B}\frac{N_{2}-1}{N_{2}}[N_{21}N_{22}\cos(2\phi_{2,2})-N_{21}N_{23}\cos(2\phi_{2,3})-N_{21}N_{24}\cos(2\phi_{2,4})\nonumber\\
&&-N_{22}N_{23}\cos2(\phi_{2,3}-\phi_{2,2})-N_{22}N_{24}\cos2(\phi_{2,4}-\phi_{2,2})+N_{23}N_{24}\cos2(\phi_{2,4}-\phi_{2,3})\nonumber\\
&&+4\sqrt{N_{21}N_{22}N_{23}N_{24}}[-\cos(\phi_{2,4}+\phi_{2,3}-\phi_{2,2})+\cos(\phi_{2,2}+\phi_{2,3}-\phi_{2,4})+\cos(\phi_{2,2}+\phi_{2,4}-\phi_{2,3})]]\nonumber\\
&&+8U_{2,B}[\sqrt{N_{11}N_{12}N_{21}N_{22}}\cos(\phi_{1,2})\cos(\phi_{2,2})-\sqrt{N_{11}N_{13}N_{21}N_{23}}\sin(\phi_{1,3})\sin(\phi_{2,3})\nonumber\\
&&-\sqrt{N_{11}N_{14}N_{21}N_{24}}\sin(\phi_{1,4})\sin(\phi_{2,4})-\sqrt{N_{12}N_{13}N_{22}N_{23}}\sin(\phi_{1,3}-\phi_{1,2})\sin(\phi_{2,3}-\phi_{2,2})\nonumber\\
&&-\sqrt{N_{12}N_{14}N_{22}N_{24}}\sin(\phi_{1,4}-\phi_{1,2})\sin(\phi_{2,4}-\phi_{2,2})+\sqrt{N_{13}N_{14}N_{22}N_{23}}
\cos(\phi_{1,4}-\phi_{1,3})\cos(\phi_{2,4}-\phi_{2,3})]\nonumber\\
&&+8U_{2,B}[\sqrt{N_{11}N_{12}N_{23}N_{24}}\cos(\phi_{1,2})\cos(\phi_{2,4}-\phi_{2,3})+\sqrt{N_{11}N_{13}N_{22}N_{24}}\sin(\phi_{1,3})\sin(\phi_{2,4}-\phi_{2,2})\nonumber\\
&&+\sqrt{N_{11}N_{14}N_{22}N_{23}}\sin(\phi_{1,4})\sin(\phi_{2,3}-\phi_{2,2})+\sqrt{N_{12}N_{13}N_{21}N_{24}}\sin(\phi_{1,3}-\phi_{1,2})\sin(\phi_{2,4})\nonumber\\
&&+\sqrt{N_{12}N_{14}N_{21}N_{23}}\sin(\phi_{1,4}-\phi_{1,2})\sin(\phi_{2,3})+\sqrt{N_{13}N_{14}N_{21}N_{22}}\cos(\phi_{1,4}-\phi_{1,3})\cos(\phi_{2,2})]\}
\sin^{4}(\tilde{\theta}/2).\nonumber
\end{eqnarray}
When the bosons are condensed at only one of the minima, for example, at ${\bf Q_{1}}$, then $N_{11}=N_{1}$ and $N_{21}=N_{2}$,
$N_{1i}=N_{2i}=0$ with $i\neq1$. As a result, it is easy to see that $<\tilde{\Psi}_{c}|H_{int}|\tilde{\Psi}_{c}>=
<\tilde{\Psi}_{f}|H_{int}|\tilde{\Psi}_{f}>=\tilde{E}_{s}$. When the bosons are only simultaneously condensed at two minima
which are time-reversal partner, for example, ${\bf Q_{1}}$ and ${\bf Q_{2}}$, then the above equation is greatly simplified,
\begin{eqnarray}
<\tilde{\Psi}_{c}|H_{int}|\tilde{\Psi}_{c}>&=&<\tilde{\Psi}_{f}|H_{int}|\tilde{\Psi}_{f}>+\{2U_{1,A}\frac{N_{1}-1}{N_{1}}
[N_{11}N_{12}\cos(2\phi_{1,2})]+2U_{1,B}\frac{N_{1}-1}{N_{1}}[N_{11}N_{12}\cos(2\phi_{1,2})]\nonumber\\
&&+[(N_{1},N_{1i},\phi_{1,i})\longleftrightarrow(N_{2},N_{2i},\phi_{2,i})]\}+8U_{2,A}[\sqrt{N_{11}N_{12}N_{21}N_{22}}\cos(\phi_{1,2})\cos(\phi_{2,2})]\nonumber\\
&&+8U_{2,B}[\sqrt{N_{11}N_{12}N_{21}N_{22}}\cos(\phi_{1,2})\cos(\phi_{2,2})]. \label{23}
\end{eqnarray}
When $U_{1}>U_{2}$, by minimizing the energy, it is found that $N_{11}$,$N_{12}$,$N_{21}$,$N_{22}$ can takes
arbitrary values if $\phi_{1,2}$, $\phi_{2,2}$ are locked to $\{(n+\frac{1}{2})\pi, n\in Z\}$, and the minimum
value of $<\tilde{\Psi}_{c}|H_{int}|\tilde{\Psi}_{c}>$ is also $\tilde{E}_{s}$. However, if the bosons are simultaneously
condensed at two minima that are not time-reversal partner, for example, ${\bf Q_{1}}$ and ${\bf Q_{3}}$, it is found that
the minimum value of $<\tilde{\Psi}_{c}|H_{int}|\tilde{\Psi}_{c}>$ is given as $\tilde{E}_{s}+4[U_{1,B}(N_{11}N_{13}+N_{21N_{23}})-2U_{2,B}\sqrt{N_{11}N_{13}N_{21}N_{23}}]\sin^{4}(\tilde{\theta}/2)$,
which is larger than $\tilde{E}_{s}$, and therefore, it is not favored in energy. Other cases can be similarly discussed
and we neglect the discussion here.

\subsection{C. Particle distribution for $\alpha>\alpha_{c}$.}

\begin{eqnarray}
|\varphi_{A\uparrow}({\bf r_{i}})|^{2}&=&n_{0}\tilde{\chi}_{1}^{2}(1+2|b_{1}||b_{3}|\cos(\frac{\pi(x_{i}+y_{i})}{a}-\varphi_{13})
+2|b_{1}||b_{5}|\cos(\frac{\pi y_{i}}{a}-\varphi_{15})+2|b_{1}||b_{7}|\cos(\frac{\pi x_{i}}{a}-\varphi_{17})\nonumber\\
&&+2|b_{3}||b_{5}|\cos(\frac{\pi x_{i}}{a}+\varphi_{35})+2|b_{3}||b_{7}|\cos(\frac{\pi y_{i}}{a}+\varphi_{37})
+2|b_{5}||b_{7}|\cos(\frac{\pi(x_{i}-y_{i})}{a}-\varphi_{57}))\xi_{1}. \nonumber\\
|\varphi_{B\downarrow}({\bf r_{i}})|^{2}&=&n_{0}\tilde{\chi}_{2}^{2}(1-2|b_{1}||b_{3}|\cos(\frac{\pi(x_{i}+y_{i})}{a}-\varphi_{13})
-2|b_{1}||b_{5}|\sin(\frac{\pi y_{i}}{a}-\varphi_{15})+2|b_{1}||b_{7}|\sin(\frac{\pi x_{i}}{a}-\varphi_{17})\nonumber\\
&&-2|b_{3}||b_{5}|\sin(\frac{\pi x_{i}}{a}+\varphi_{35})+2|b_{3}||b_{7}|\sin(\frac{\pi y_{i}}{a}+\varphi_{37})
-2|b_{5}||b_{7}|\cos(\frac{\pi(x_{i}-y_{i})}{a}-\varphi_{57})\xi_{2}. \nonumber\\
|\varphi_{B\uparrow}({\bf r_{i}})|^{2}&=&n_{0}\tilde{\chi}_{1}^{2}(1-2|b_{2}||b_{4}|\cos(\frac{\pi(x_{i}+y_{i})}{a}-\varphi_{24})
+2|b_{2}||b_{6}|\sin(\frac{\pi y_{i}}{a}-\varphi_{26})-2|b_{2}||b_{8}|\sin(\frac{\pi x_{i}}{a}-\varphi_{28})\nonumber\\
&&+2|b_{4}||b_{6}|\sin(\frac{\pi x_{i}}{a}+\varphi_{46})-2|b_{4}||b_{8}|\sin(\frac{\pi y_{i}}{a}+\varphi_{48})
-2|b_{6}||b_{8}|\cos(\frac{\pi(x_{i}-y_{i})}{a}-\varphi_{68}))\xi_{2}. \nonumber\\
|\varphi_{A\downarrow}({\bf r_{i}})|^{2}&=&n_{0}\tilde{\chi}_{1}^{2}(1+2|b_{2}||b_{4}|\cos(\frac{\pi(x_{i}+y_{i})}{a}-\varphi_{24})
+2|b_{2}||b_{6}|\cos(\frac{\pi y_{i}}{a}-\varphi_{26})+2|b_{2}||b_{8}|\cos(\frac{\pi x_{i}}{a}-\varphi_{28})\nonumber\\
&&+2|b_{4}||b_{6}|\cos(\frac{\pi x_{i}}{a}+\varphi_{46})+2|b_{2}||b_{4}|\cos(\frac{\pi y_{i}}{a}+\varphi_{48})
+2|b_{6}||b_{8}|\cos(\frac{\pi(x_{i}-y_{i})}{a}-\varphi_{68}))\xi_{1}. \nonumber\\
\end{eqnarray}
$|b_{i\in odd}|$ correspond to $\hat{\alpha}_{1}$, while $|b_{i\in even}|$ correspond to $\hat{\alpha}_{2}$.

When only one of $|b_{i\in odd}|$ and one of $|b_{i\in even}|$ are nonzero, it is direct to see that
$|\varphi_{A\uparrow}({\bf r_{i}})|^{2}=|\varphi_{A\downarrow}({\bf r_{i}})|^{2}=n_{0}\tilde{\chi}_{1}^{2}$,
$|\varphi_{B\uparrow}({\bf r_{i}})|^{2}=|\varphi_{B\downarrow}({\bf r_{i}})|^{2}=n_{0}\tilde{\chi}_{2}^{2}$,
which corresponds to a spin-balanced condensate. When the bosons are condensed at two minima which are
time-reversal partner, for example, ${\bf Q_{1}}$ and ${\bf Q_{2}}$, then only $b_{1,2,3,4}$ are nonzero.
When $\varphi_{13}$ (equivalent to $\phi_{1,2}$) and $\varphi_{24}$ (equivalent to $\phi_{2,2}$) are locked
to $\{(n+\frac{1}{2})\pi, n\in Z\}$, it is easy to see that the spin configuration is the same as the former
single minimum occupied case, with $|\varphi_{A\uparrow}({\bf r_{i}})|^{2}=|\varphi_{A\downarrow}({\bf r_{i}})|^{2}=n_{0}\tilde{\chi}_{1}^{2}$,
$|\varphi_{B\uparrow}({\bf r_{i}})|^{2}=|\varphi_{B\downarrow}({\bf r_{i}})|^{2}=n_{0}\tilde{\chi}_{2}^{2}$.
A discussion of the spin configuration corresponding to other possible ground states is similar and we neglect
it here.

\end{widetext}

\end{document}